\documentclass{article}
\usepackage[margin=1in]{geometry}
\usepackage{graphicx}
\usepackage[square]{natbib}
\usepackage[hidelinks]{hyperref}
\usepackage{amsmath,amssymb,amsfonts,amsthm,mathrsfs}
\usepackage{booktabs,longtable,multirow,array,makecell,threeparttable}
\usepackage{subcaption,float}
\usepackage[table,xcdraw]{xcolor}
\usepackage{bm}
\usepackage{authblk}

\newcommand{\indep}{\mathrel{\perp\!\!\!\perp}}
\newcommand{\PM}{PM$_{2.5}$}
\newcommand{\mugm}{\ensuremath{\mu\mathrm{g}/\mathrm{m}^3}}
\newcommand{\degree}{\ensuremath{^\circ}}
\newcommand{\bmhead}[1]{\par\medskip\noindent\textbf{#1}\par}

\begin{document}

\title{\textbf{Evaluating differential air quality impacts of prescribed fire and wildfire in the United States}}
\author[1]{Ting-Hsuan Chang}
\author[2,3]{Minghao Qiu}
\author[4]{Yaguang Wei}
\author[1]{Xiao Wu\thanks{Corresponding author: xw2892@cumc.columbia.edu}}
\affil[1]{Department of Biostatistics, Columbia University Mailman School of Public Health}
\affil[2]{School of Marine and Atmospheric Sciences, Stony Brook University}
\affil[3]{Program in Public Health, Stony Brook University}
\affil[4]{Department of Environmental Medicine, Icahn School of Medicine at Mount Sinai}
\date{September 11, 2026}

\maketitle

\section*{Abstract}

Increasing wildfire activities in the United States have generated far-reaching effects on air pollution and human health. While prescribed fires are used as a preventive strategy to reduce wildfire risks, the health impacts of their smoke remain uncertain. Small-scale case studies indicate that prescribed fires generally produce lower pollutant emissions than uncontrolled wildfires, potentially benefiting air quality. However, the extent to which this translates into reduced population-level smoke exposure remains uncertain due to a lack of robust, large-scale quantitative evidence. Moreover, the factors driving differences in air quality impacts between prescribed fires and wildfires were underexplored. Using fire data in California, Florida, and Georgia (2006-2020) and a causal forest approach, we quantified and compared the impacts of prescribed fires and wildfires on air quality, assessed their differences in population-level smoke exposure, and identified key factors influencing their smoke exposure levels. On a \textit{per square-kilometer} basis, prescribed fires resulted in over six times less population-level smoke exposure compared to wildfires. Additionally, prescribed fires were associated with lower smoke exposure in regions with higher relative humidity and precipitation. This work could enable the broad risk-benefit assessment to guide the judicious use of prescribed fires and optimize public health outcomes while mitigating wildfire threats.

\medskip
\noindent\textbf{Keywords:} Exposure, Fire Management, Particles, Pollution

\section{Introduction}

More acres have burned year over year in many regions of the United States (US) \citep{RN19}, attributed to changing climate and legacy forest management failure \citep{RN2,RN26}, increasing the frequency and severity of wildfires along with their emission. On the one hand, rising temperatures cause more severe droughts and longer fire seasons \citep{RN27}, drying out vegetation and making land more susceptible to wildfires. Additionally, warmer weather increases the frequency of lightning strikes that can spark fires \citep{RN28}. On the other hand, legacy land management policies have contributed to the increased risk of large and severe wildfires in the long term \citep{RN2}. Widespread fire suppression has led to landscapes with excessive fuel accumulation that greatly increases the risk and severity of future fires \citep{turner2003surprises}. The post-fire removal of burned trees in the name of restoration has hindered the recovery of ecosystem functions and biodiversity \citep{moritz2014learning}. Human-grown plantations of even-aged, single-species vegetation have increased fire spread and compromised the potential for natural regeneration \citep{taylor2014nonlinear}.

Wildfires create public health concerns beyond the boundaries of fires. Wildfires are a significant source of air pollution, emitting a combination of particles and gases that adversely affect health \citep{burke2023contribution,RN30}. The main component of wildfire smoke is particulate matter pollution, especially fine particulate matter with a diameter less than 2.5 microns  (\PM), which represents a major public health threat \citep{RN31}. Wildfires contributed to a sharp increase in the level of \PM\ in many states \citep{RN21} and led to stagnation or even reversal of the otherwise declining trend in ambient \PM\ over the last two decades due to the Clean Air Act \citep{burke2023contribution}. A 2024 report found that 131.2  million of Americans were living in areas with unhealthy levels of air pollution, with wildfire smoke identified as an important contributing factor \citep{RN21}. The impact of wildfire smoke extends well beyond the immediate vicinity of the fires: most of the smoke \PM\ comes from sources outside the local jurisdiction, with 87\% from fires in other counties and 60\% from out-of-state fires \citep{RN28}. Wildfire smoke has been associated with adverse health outcomes, including respiratory, cardiovascular, and nervous diseases. Adding to the burden is enormous economic cost: wildfire-related deaths and healthcare costs in the US now total tens of billions annually \citep{RN22} and are projected to be substantially increasing under future climate scenarios \citep{qiu2024wildfire}. A 2023 systematic review provided high-confidence evidence linking wildfire smoke exposure to increased all-cause mortality and respiratory morbidity, while highlighting emerging associations with cognitive disorders and other disease risks \citep{RN4}. 

The substantial impacts of wildfires and wildfire smoke drive scientists, policymakers, and at-risk communities to seek effective tools to reduce wildfire risk \citep{RN2}. Planned and intentionally set controlled burns, known as prescribed fires, have been used to reduce the risk of unwanted wildfires \citep{RN34}. Recently, prescribed fire has attracted increasing policy discussions due to its cost-effectiveness \citep{RN34} and is often viewed as a key pillar of future wildfire policies. For example, California has proposed treating up to 1 million acres of land annually by 2025, including using prescribed fires to reduce fuel load \citep{forest2021california}. Prescribed fires reduce the fuel available for unwanted wildfires, increase landscape heterogeneity, improve resilience to fire and drought-related disturbance, and increase soil moisture and runoff \citep{RN29}. In contrast, if these measures are not adequately funded or implemented, forests can become overgrown and dense, allowing wildfires to spread more quickly and become more intense.  Evidence has demonstrated the effectiveness of prescribed fire in reducing wildfire risk, particularly for high-intensity wildfires \citep{thompson2007reburn,parks2014previous,RN2}.

Despite these compelling benefits, broader implementation of prescribed fires faces significant obstacles.  Smoke from fires remains a major concern, often exceeding designated control boundaries and escaping the control of fire managers \citep{RN36}. Although a 2024 study indicated that prescribed fires led to a net reduction in smoke emissions \citep{kelp2024efficacy}, it is unclear whether this reduction translates to lessened population exposure to smoke-related air pollutants. 
Additionally, it is unclear how factors influencing fire behavior—weather, topography, and vegetation (fuel)—differ between prescribed fires and wildfires, and how these differences may exacerbate or mitigate smoke impacts. Given the desire to strategically choose the locations and timing of prescribed fires, identifying these factors is critical for informing future prescribed fire planning to minimize public health risks while reducing wildfire threats.

Our study aimed to provide large-scale quantitative evidence on air pollution exposure from wildland fires, including both wildfires and prescribed fires. We compare the relative severity of population-level smoke exposure from these two types of fires and identify key factors contributing to the variability in smoke exposure. To achieve this, we integrate modern statistical techniques with geospatial data pipelines. Causal inference methods are particularly suitable in isolating the impact of fire on smoke exposure while accounting for other influencing factors \citep{imbens2015causal}. These methods provide an interpretable framework for comparing outcomes induced by different interventions (i.e., fires in our study). Specifically, we use a causal forest approach \citep{RN37}, which allows a straightforward interpretation of which factors contribute significantly to the heterogeneity of fire-induced smoke exposure \citep{yadlowsky2025evaluating}. Through this approach, we assess how smoke exposure varies by fire type and other influencing factors, accounting for environmental and geographical differences across fire areas and states. This approach captures localized fire-smoke relationships, offering a flexible and data-driven perspective on the relationship between wildland fires and smoke exposure.

Further, we create data pipelines that bring together information on topography, weather, and vegetation before the ignition and the fire duration from all recorded fires 2006-2020 in California (CA), Florida (FL), and Georgia (GA), three states with distinct landscapes and climates. We link each fire footprint with its area- and population-level smoke \PM\ exposure \citep{wen2023quantifying}. 
By considering differences in environmental and geographical conditions across three states,  we could capture a broad spectrum of fire behavior and gain insight into how terrain, weather, and vegetation shape the air quality impacts of fires.

\section{Results}

\subsection{Fire Events}
\begin{figure}[t]
\centering
\includegraphics[width=16cm,height=8.5cm]{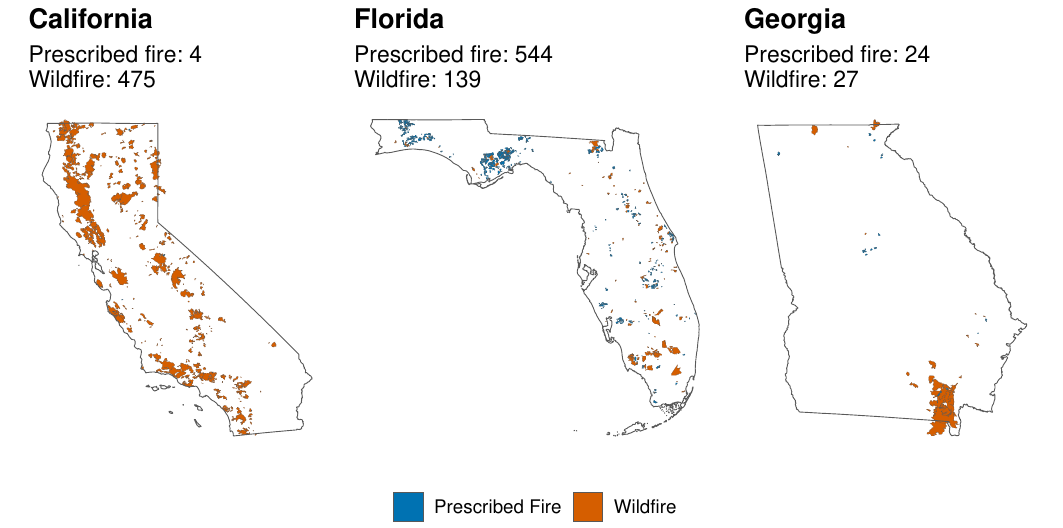}
\caption{Fire distributions and the specific fire types by state from 2006 to 2020.}\label{fig:state_map}
\end{figure}

Our study included 1,229 fire events: 579 (47.1\%) prescribed fires and 650 (52.9\%) wildfires, derived from multi-sourced fire and geographical datasets (see \textit{Methods}). Fire types were identified by systems of fire record and report provided by state and federal agencies/organizations \citep{eidenshink2007project}. These fires were ignited between 2006 and 2020 in CA, FL, and GA. Fig.~\ref{fig:state_map} shows the fire distributions and fire types by each state. The average duration for all fires was 9.75 days (median = 7, min = 1, max = 67). The average duration was 7.71 days (median = 7, min = 1, max = 29) for prescribed fires and 11.6 days (median = 8, min = 1, max = 67) for wildfires. Wildfires were also, on average, larger in size than prescribed fires -- the mean area size of wildfires was 96.4 km$^2$ (median = 21.4, min = 2.26, max = 4325), while the mean area size of prescribed fires was 8.43 km$^2$ (median = 6.63, min = 2.08, max = 75.7).

We compared prescribed fires and wildfires in a wide range of topography, weather, and vegetation variables before or during their ignition date in Table~\ref{tab:base}. The environmental conditions in which prescribed fires and wildfires occurred were quite different. For example, wildfires tended to occur in environments with higher elevation, greater vapor pressure deficit, lower humidity, and less forest coverage compared to prescribed fires.

\begin{center}
\onecolumn
\begin{longtable}{llll}
\caption{Comparison of topography, weather, and vegetation before or during ignition between prescribed fires and wildfires}
    \label{tab:base}\\
\toprule
  & Prescribed Fire & Wildfire & Overall\\
\midrule
 & (N=579) & (N=650) & (N=1229)\\
\addlinespace[0.05em]
\multicolumn{4}{l}{\textbf{Elevation (m)}}\\
\hspace{1em}Median [Min, Max] & 29.4 [2.90, 2040] & 706 [3.03, 2690] & 51.7 [2.90, 2690]\\
\addlinespace[0.05em]
\multicolumn{4}{l}{\textbf{Slope (\degree)}}\\
\hspace{1em}Median [Min, Max] & 0.384 [0.0482, 15.9] & 8.02 [0.0124, 26.6] & 0.748 [0.0124, 26.6]\\
\addlinespace[0.05em]
\multicolumn{4}{l}{\textbf{Aspect sine (\degree)}}\\
\hspace{1em}Median [Min, Max] & -0.00139 [-1360, 0.879] & -0.00829 [-6000, 0.909] & -0.00507 [-6000, 0.909]\\
\addlinespace[0.05em]
\multicolumn{4}{l}{\textbf{Aspect cosine (\degree)}}\\
\hspace{1em}Median [Min, Max] & -0.0853 [-1360, 0.962] & -0.0916 [-6000, 0.944] & -0.0892 [-6000, 0.962]\\
\addlinespace[0.05em]
\multicolumn{4}{l}{\textbf{Precipitation (mm)}}\\
\hspace{1em}Median [Min, Max] & 0 [0, 38.9] & 0 [0, 31.7] & 0 [0, 38.9]\\
\addlinespace[0.05em]
\multicolumn{4}{l}{\textbf{Wind direction (\degree)}}\\
\hspace{1em}Median [Min, Max] & 225 [2.00, 360] & 218 [1.00, 357] & 221 [1.00, 360]\\
\addlinespace[0.05em]
\multicolumn{4}{l}{\textbf{Wind velocity (m/s)}}\\
\hspace{1em}Median [Min, Max] & 3.50 [0.800, 10.5] & 3.00 [0.600, 11.0] & 3.20 [0.600, 11.0]\\
\addlinespace[0.05em]
\multicolumn{4}{l}{\textbf{Vapor pressure deficit (kPa)}}\\
\hspace{1em}Median [Min, Max] & 0.830 [0, 2.15] & 1.97 [0.110, 5.01] & 1.23 [0, 5.01]\\
\addlinespace[0.05em]
\multicolumn{4}{l}{\textbf{Min. temperature (K)}}\\
\hspace{1em}Median [Min, Max] & 282 [263, 298] & 288 [270, 302] & 286 [263, 302]\\
\addlinespace[0.05em]
\multicolumn{4}{l}{\textbf{Max. temperature (K)}}\\
\hspace{1em}Median [Min, Max] & 297 [277, 312] & 304 [283, 315] & 301 [277, 315]\\
\addlinespace[0.05em]
\multicolumn{4}{l}{\textbf{Min. relative humidity (\%)}}\\
\hspace{1em}Median [Min, Max] & 36.2 [16.2, 69.7] & 18.8 [1.00, 78.9] & 28.5 [1.00, 78.9]\\
\addlinespace[0.05em]
\multicolumn{4}{l}{\textbf{Max. relative humidity (\%)}}\\
\hspace{1em}Median [Min, Max] & 92.5 [38.8, 100] & 56.9 [11.4, 100] & 84.1 [11.4, 100]\\
\addlinespace[0.05em]
\multicolumn{4}{l}{\textbf{Forest coverage (\%)}}\\
\hspace{1em}Median [Min, Max] & 44.0 [0, 96.4] & 16.4 [0, 99.2] & 32.6 [0, 99.2]\\
\addlinespace[0.05em]
\multicolumn{4}{l}{\textbf{Shrubland coverage (\%)}}\\
\hspace{1em}Median [Min, Max] & 3.82 [0, 87.2] & 13.4 [0, 99.5] & 7.96 [0, 99.5]\\
\addlinespace[0.05em]
\multicolumn{4}{l}{\textbf{Herbaceous coverage (\%)}}\\
\hspace{1em}Median [Min, Max] & 0.379 [0, 82.1] & 5.44 [0, 99.6] & 1.38 [0, 99.6]\\
\bottomrule
\end{longtable}
\end{center}

\subsection{Smoke Exposure}
Smoke exposure was calculated as the sum of daily exposure to smoke \PM\ for all downwind locations over the duration of a fire \citep{wen2023quantifying}. Specifically, the air parcel trajectories of individual fires were overlayed with population data to compute the cumulative smoke \PM\ experienced by the affected population in the contiguous US. Since wildfires were, on average, much larger in size than prescribed fires, we normalized smoke exposure by burned area (per km$^2$) to account for differences in fire size. We refer to the population-weighted smoke exposure (in person-\mugm), divided by fire size (km$^2$), as \textit{population smoke exposure per $km^2$}. The cumulative concentration of smoke (in \mugm) without accounting for the population exposed, divided by fire size, is referred to as \textit{area smoke exposure per $km^2$}.

Table \ref{tab:estimates} shows the estimated impacts of fire on population and area smoke exposure per km$^2$ by each fire type. Overall, we found that wildfires led to notably higher smoke exposure than prescribed fires of the same size. The average population smoke exposure per km$^2$ for prescribed fires was $2.4 \times 10^5$ (95\% CI: [$8.5 \times 10^4$, $4.0 \times 10^5$]) person-\mugm. For wildfires, the average population smoke exposure was $1.6 \times 10^6$ (95\% CI: [$1.0 \times 10^6$, $2.1 \times 10^6$]) person-\mugm. On a per-km$^2$ basis, wildfires led to more than six times the population smoke exposure than prescribed fires. Given that wildfires are generally larger in size than prescribed fires, the overall population smoke exposure, accounting for fire size, is expected to be even greater for wildfires than for typical prescribed fires. Indeed, wildfires led to more than 21 times the population smoke exposure than prescribed fires if the per-km$^2$ estimates were multiplied by the corresponding median fire size. 

\begin{table}[ht] 
\centering
\caption{Smoke exposure estimates per km$^2$ fire size}
\begin{tabular}[h]{lcc}
\toprule
  & \thead{Prescribed fires} & \thead{Wildfires} \\
\midrule
\addlinespace[0.3em]
\multicolumn{3}{l}{Population smoke exposure (person-\mugm)}\\
\hspace{1em}Mean & $2.4 \times 10^5$ & $1.6 \times 10^6$ \\
\hspace{1em}95\% CI & [$8.5 \times 10^4$,\quad$4.0 \times 10^5$] & [$1.0 \times 10^6$,\quad$2.1 \times 10^6$] \\
\addlinespace[0.3em]
\multicolumn{3}{l}{Area smoke exposure (\mugm)}\\
\hspace{1em}Mean & $44.3$ & $615.8$ \\
\hspace{1em}95\% CI & $[26.2,\quad62.4]$ & $[432.4,\quad799.2]$ \\
\bottomrule
\end{tabular}
\label{tab:estimates}
\end{table}

\subsection{Factors Moderating Fire Impacts on Smoke}
\begin{figure}[t]
\centering
\includegraphics[width=16cm,height=6cm]{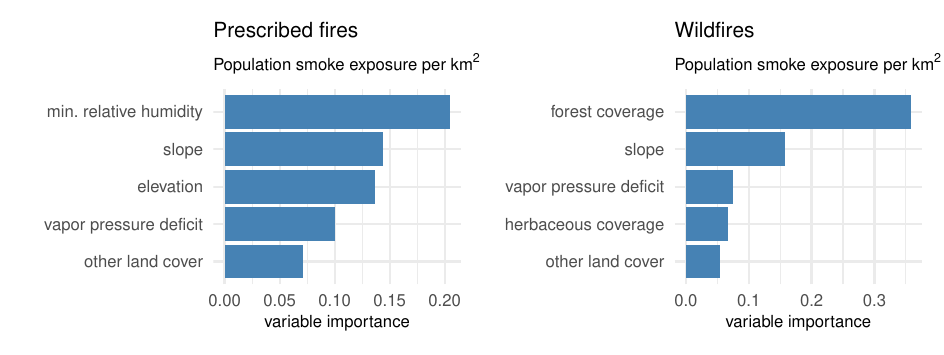}
\caption{Variable importance scores in moderating the population smoke exposure (log person-\mugm) per km$^2$ of prescribed fires (left) and wildfires (right).}\label{fig:varimp_pop}
\end{figure}

Topography, weather, and vegetation variables, shown in Table \ref{tab:base}, along with the burn severity, may potentially moderate (i.e., worsen or lessen) the smoke exposure from prescribed fires or wildfires. Their moderating effects were captured using a variable importance score, which reflects a variable's contribution to the overall heterogeneity in smoke exposure (see \textit{Methods}). Fig.~\ref{fig:varimp_pop} shows the top five important variables for population smoke exposure per km$^2$ on the logarithmic scale. Variables such as minimum relative humidity, slope, and elevation appeared to play a critical role in moderating population smoke exposure from prescribed fires (left of Fig.~\ref{fig:varimp_pop}), while forest coverage and slope appeared as key factors in moderating population smoke exposure from wildfires (right of Fig.~\ref{fig:varimp_pop}). The importance scores of the top five variables for moderating area smoke exposure showed similar results overall (see the Supplementary information).

\begin{figure}[t]
\centering
\includegraphics[width=16cm,height=10cm]{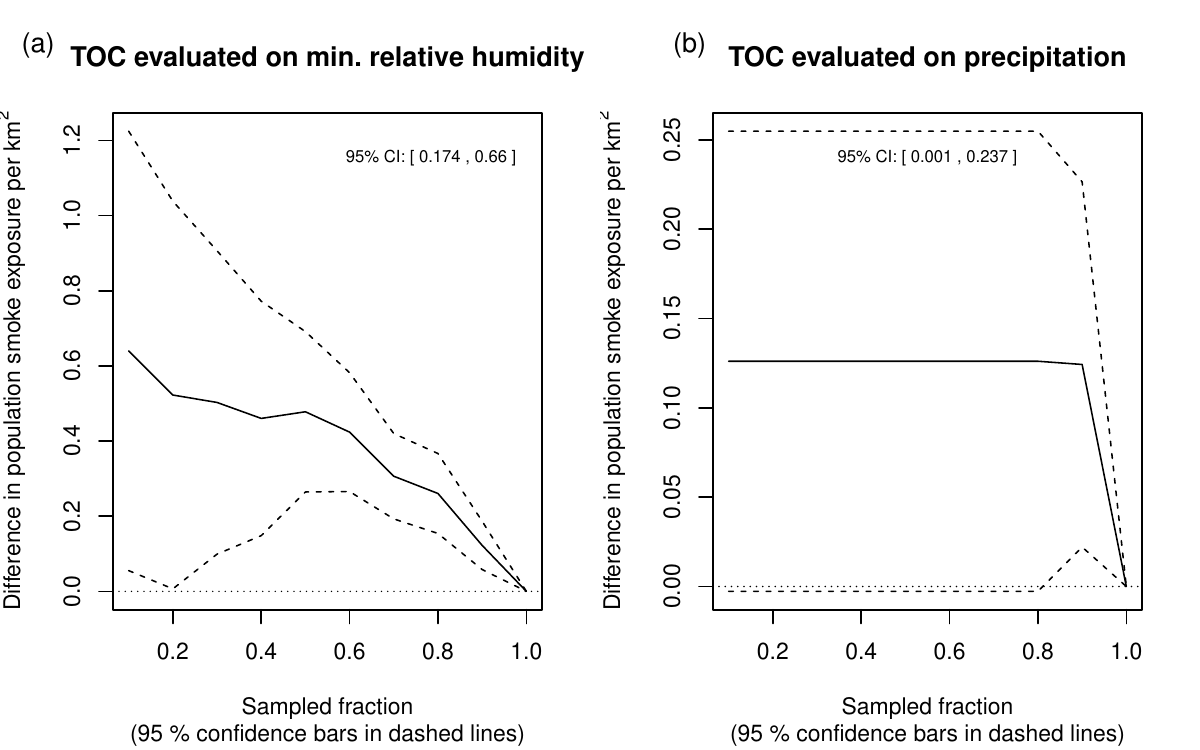}
\caption{Targeting Operator Characteristic (TOC) curves for the variation in population smoke exposure (log person-\mugm) per km$^2$ from prescribed fires attributed to each variable. The y-axis shows how smoke exposure varies in regions with different values of a specific variable (e.g., minimal relative humidity and precipitation) compared to the overall average. A downward-trending curve indicates that lower values of the evaluated variable are associated with higher smoke exposure. The 95\% confidence interval (CI) of the area under the TOC curve (AUTOC) is displayed in each plot. Only statistically significant results are shown, while TOC curves for other variables are included in the Supplementary information.}\label{fig:pres_pop}
\end{figure}

Targeting Operator Characteristic (TOC; \citep{yadlowsky2025evaluating}) curves (e.g., Fig.~\ref{fig:pres_pop}) were used to assess the direction of moderating effect of each variable on smoke exposure. In each panel, the y-axis shows the difference in smoke exposure per km$^2$ (log scale) between (1) the top $q$-th fraction of the sample with the smallest values of the evaluated variable and (2) the entire sample. A downward-trending curve indicates that lower values of the evaluated variable are associated with higher smoke exposure. 

For prescribed fires, population smoke exposure tended to be lower in areas with higher minimum relative humidity (Fig.~\ref{fig:pres_pop}(a)) and in areas experiencing very high levels of precipitation (Fig.~\ref{fig:pres_pop}(b)). For wildfires, lower population smoke exposure was linked to areas with less forest coverage, gentler slopes, lower vapor pressure deficits, lower elevations, and higher minimum relative humidity (see Fig. B4). We did not find statistically significant moderating effects for other variables among the ten highest-ranked in variable importance scores. Higher humidity level appeared to mitigate the population smoke exposure from both prescribed fires and wildfires. Prescribed fires were generally administered in more humid environments than wildfire sites (Table \ref{tab:base}), which may partly explain their lower smoke impact. TOC plots for the ten variables with the highest importance scores are provided in the Supplementary Information. The findings for area smoke exposure are largely consistent with those for population smoke exposure. 



\subsection{Comparison between Prescribed Fires and Wildfires}
Among areas with a likelihood of experiencing either fire type, we compared the expected smoke exposure per km$^2$ under a prescribed fire versus a wildfire. Given the substantial environmental and geographical differences between prescribed fire and wildfire sites, this comparison accounted for variations in topography, weather, and vegetation, as well as potential state-level baseline differences between the two fire types. Overall, prescribed fires had much lower impacts on smoke exposure relative to wildfires. Prescribed fires were estimated to produce $1.3 \times 10^6$ person-\mugm\ (95\% CI: [$-2.0 \times 10^6, -5.2 \times 10^5$]) less in population smoke exposure than wildfires, and $212.6$ \mugm\  (95\% CI: [$-350.4, -74.8$]) less in area smoke exposure (Table~\ref{tab:pres_vs_wild}). Table~\ref{tab:pres_vs_wild} presents a counterfactual comparison where topography, weather, and vegetation conditions are held constant. Compared to the results in Table~\ref{tab:estimates}, the observed differences are smaller, suggesting that variations in these factors account for a portion of the differences in smoke exposure between prescribed fires and wildfires. However, additional unobserved factors related to fuel load, fire combustion efficiency, and atmospheric transport, might also influence smoke exposure levels.

\begin{table}[ht] 
\centering
\caption{Smoke exposure difference per km$^2$ fire size}
\begin{tabular}[t]{lcc}
\toprule
  & \thead{Population smoke exposure\\(person-\mugm)} 
  & \thead{Area smoke exposure\\(\mugm)} \\
\midrule
\addlinespace[0.3em]
\hspace{1em}Difference & $-1.3 \times 10^6$ & $-212.6$ \\
\hspace{1em}95\% CI & [$-2.0 \times 10^6$,\quad$-5.2 \times 10^5$] & [$-350.4$,\quad$-74.8$] \\
\bottomrule
\end{tabular}
\label{tab:pres_vs_wild}
\end{table}

\section{Discussion}

In this study, we quantified the differential effects of wildfires and prescribed fires on air quality using statistical learning tools for causal inference and satellite data on fire, fire smoke, and fire behavior-related conditions in CA, FL, and GA. The main findings are twofold: (1) It confirmed, on a large scale, that prescribed fires result in substantially less smoke exposure for human populations compared to wildfires, even after adjusting for important determinants of fire behavior. (2) Environmental factors, particularly higher relative humidity and precipitation, were found to reduce population-level smoke exposure from prescribed fires.

The findings of this study have important policy implications related to air quality and wildland fire policies. Since 2016, ambient \PM\ concentrations have been rising across many regions of the US, driven in part by wildfire smoke, demanding policy responses \citep{burke2023contribution}. However, there is a policy dilemma under the Clean Air Act: while wildfire smoke as a natural source is explicitly exempt from both local and transboundary attainment rules for air quality, prescribed fire is regulated under the Act as it is considered an anthropogenic emission source \citep{engel2013perverse}. Our results could support revisions of key air quality regulations, making air quality exemptions for prescribed fires conditional on efforts to reduce wildfire risk, as well as easing the permitting barriers for prescribed fire smoke emissions. Our study also provides a foundation for future wildland fire policies that prioritize reducing population exposure rather than focusing solely on emission reduction. The US Forest Service has proposed large-scale investment in prescribed fire programs to reduce the risk of extreme wildfires \citep{usda2022confronting}. In addition to supporting the expansion of prescribed fires, our study provides a framework for considering the timing and placement of prescribed fires to maximize health benefits through reductions in smoke exposure.

The strengths of this study include: first, our study harmonized large-scale spatial-temporal fire and smoke-related data from various satellite remote sensing tools. This data integration leverages the unique strengths of each dataset, offering deeper insights into fire dynamics, environmental impacts, and the development of potential mitigation strategies. Second, we applied the causal forest method, which enabled an interpretable characterization of the effect heterogeneity in smoke exposure. We also formally defined the differential impact of prescribed fires versus wildfires on smoke exposure while controlling for potential confounders at both the area and state levels. Third, we made all code for processing and analyzing the data publicly accessible to ensure the reproducibility and transparency of our results.

The limitations of this study include: first, prescribed fire data sources were limited in their resolution and coverage. We only accessed relatively large fires through the MTBS dataset, while many prescribed fires may occur on smaller, more localized scales. Therefore, our results should be interpreted in the context of relatively large prescribed fires, although we anticipate that smaller prescribed fires would have even less impact on air quality. Future research could expand the analysis to include both large and small prescribed fires if more comprehensive satellite-based observations become available. Second, cross-linking fire boundary and smoke data across multiple data sources presents challenges. Spatial and temporal misalignments of fire boundary shapes can result in mismatches or missing data in the study. Nevertheless, we employed sophisticated spatial aggregation techniques using spatial statistical software. Third, causal forests relied on a set of assumptions to identify and estimate the impact of wildland fires on smoke exposure \citep{zhao2022selective}. Unless all assumptions are satisfied, unbiased estimation of intervention effects is not guaranteed. In particular, the assumption of no unmeasured confounding is crucial to identifying the effect of a fire intervention from observational studies, but it is often unverifiable practically. We included state indicators in addition to all measured confounders in our model to partially address potential residual confounding from unmeasured state-level confounding factors. Finally, the smoke \PM\ concentration data used in our work is derived from a machine learning model that combines satellite information and meteorology to predict surface pollution anomalies \citep{childs2022daily}. Therefore, it is possible that the smoke \PM\ estimates do not capture all the air quality impacts, especially for the smaller fires.  

Overall, our study contributes to a deeper understanding of the impacts of prescribed fires and wildfires on air pollution at the population level, highlighting that the strategic use of prescribed fires can significantly reduce air pollution exposure to the human population. These findings underscore the importance of increased investment in prescribed fires as a proactive strategy to mitigate the growing wildfire crisis in the US, while also guiding the development of equitable mitigation policies to address wildfire-related air pollution and its broader societal impacts.

\section{Methods}

\subsection{Data}
We leveraged two publicly available fire datasets: (1) The Monitoring Trends in Burn Severity (MTBS) Burned Areas Boundaries Dataset \citep{eidenshink2007project}, which contains fire boundaries from large prescribed fires and wildfires across the US from 1984 to the present. The MTBS program includes all fires of 1,000 acres or more in the western US and 500 acres or more in the eastern US \citep{eidenshink2007project}. We focused on fires that took place in three states -- CA, FL, and GA -- from 2006 to 2020, including 4,552 observations (58\% prescribed fires, 26\% wildfires, 16\% unknown). (2) The GlobFire v3 dataset \citep{artes2019global}, which includes fire boundaries used for linking fire to smoke data \citep{wen2023quantifying}. We selected CA, FL, and GA for their diverse geographical and climatic conditions, varied fire management practices, and ecosystem variability \citep{huang2018burned}. CA is known for its Mediterranean climate and significant wildfire activity, particularly in its forests and shrublands, making it an ideal location to study the impacts of large-scale wildfires. FL offers a contrasting environment with its humid subtropical climate and frequent prescribed fires, especially in its pine flatwoods and marshlands. GA provides diversity with its mix of temperate and subtropical climates, experiencing both wildfires and prescribed fires in its varied ecosystems. We joined MTBS and GlobFire datasets based on spatial and temporal alignment. Specifically, each MTBS observation was matched to GlobFire observation(s) with overlapped polygon (fire boundary) and with ignition dates within 30 days of the MTBS observation. For non-unique matches (that is, multiple GlobFire observations matched to the same MTBS observation), we only retained the matched pair that shares the largest overlapping area. After matching, we excluded pairs where the overlapping percentage ([overlapping area / MTBS fire area] $\times$ 100\%) is below the 1st quartile (21.5\%). This process yielded 2,191 observations (38.4\% wildfire, 48.8\% prescribed, 12.8\% unknown).

Burn severity data was sourced from the MTBS dataset at \href{https://www.mtbs.gov}{https://www.mtbs.gov} \citep{eidenshink2007project}, which provides burn severity classes (Unburned-Low, Low, Moderate, High, Missing) for every 30-m$^2$ pixels within the boundaries of each fire. Using all pixels whose centroids fall within the fire boundary, the burn severity of each fire was assigned based on the predominant burn severity class within the boundary. In our study, 5.2\% of prescribed fires are classified as moderate or high, compared to 28.0\% of wildfires.

Topographic variables (elevation, slope, aspect sine, and aspect cosine) were obtained from the EarthEnv data repository at \href{http://www.earthenv.org/topography}{http://www.earthenv.org/topography} \citep{amatulli2018suite}. Elevation refers to the vertical height of a point on the earth's surface, while slope indicates the steepness of a surface. Aspect sine and aspect cosine are used to represent the directional component of a slope: aspect sine represents the east-west orientation and aspect cosine represents the north-south orientation, both derived from the slope angle. For each fire, the average of pixel values within the fire boundary was calculated for each topographic variable.

Vegetation land cover information was obtained from the National Land Cover Database (NLCD), available at \href{https://www.mrlc.gov/data}{https://www.mrlc.gov/data}, which includes land cover data for specific years over the past two decades. We used data released in the years 2006, 2008, 2011, 2013, 2016, and 2019. For each fire, we used the most recent data released prior to or in the same year as the fire's occurrence. Land cover classes in our study include ``Forest," ``Shrubland," and ``Herbaceous," while remaining NLCD land cover classes (``Developed," ``Barren," ``Planted/Cultivated," ``Wetlands") were grouped as ``Other." For each fire, we extracted land cover classes, represented by pixel values, within the fire boundary and calculated the percentage of each land cover class within the boundary.

Weather variables (precipitation, wind direction, wind velocity, pressure, minimum and maximum temperature, minimum and maximum relative humidity) were obtained from the GRIDMET: University of Idaho Gridded Surface Meteorological Dataset \citep{abatzoglou2013development}. For each fire, we used the daily average of each weather variable from the day prior to the fire ignition date (as recorded in MTBS) to avoid post-intervention bias (``intervention" refers to the occurrence of fire). We converted the climate raster data into polygons and, for each fire, used the weather values from the polygon with the largest overlap with the fire boundary.

Data for fire smoke exposure was obtained from Wen et al. \citep{wen2023quantifying}, available at \href{https://github.com/jeffwen/smoke_linking_public}{https://github.com/jeffwen/smoke\_linking\_public}. We referred to the sum of the population exposed to each \mugm\ of smoke on each day over the fire duration as \emph{population smoke exposure}, and the cumulative \mugm\ of smoke over the fire duration as \emph{area smoke exposure}. We normalized these smoke exposure quantities by the corresponding fire size (in km$^2$). Hence, ``per km$^2$" in this paper refers to one square kilometer of the fire boundary. 

\begin{figure}[t]
\centering
\includegraphics[width=14cm,height=16cm]{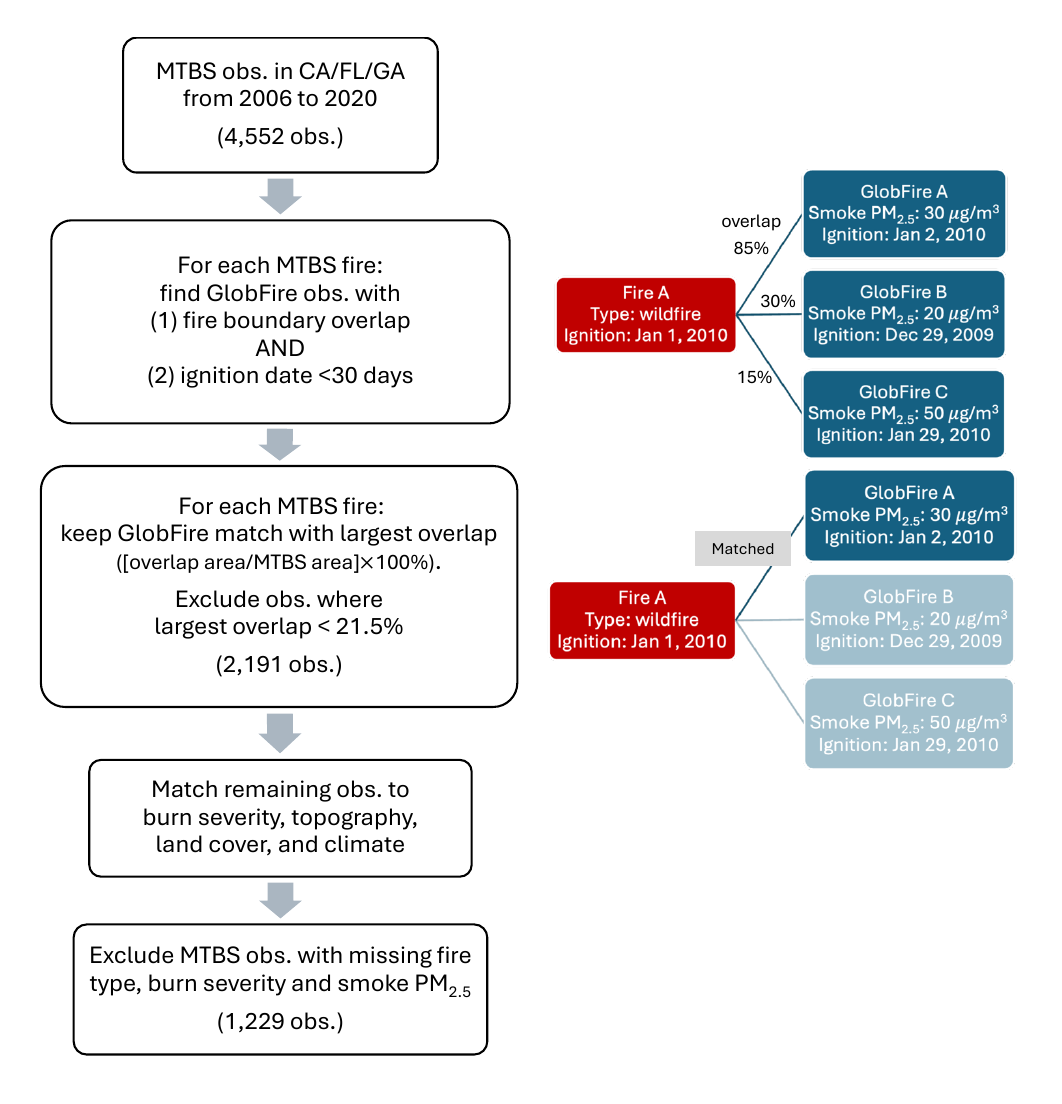}
\caption{Illustration of data merging process and sample selection. The left side of the figure shows the steps for merging different datasets. The right side illustrates how smoke exposure information is obtained for each MTBS observation (red) by matching it to a GlobFire observation (blue).}\label{fig:data_join}
\end{figure}

Our study included 1,229 complete cases. From the original sample of 2,191 fires, 822 (462 prescribed fires, 178 wildfires, and 182 of unknown type) were excluded due to missing smoke exposure data. Fires of unknown type and burn severity were also excluded. Fig.~\ref{fig:data_join} illustrates our data merging and selection process.

\subsection{Notations of statistical methods}
Let $i=1,...,n$ index the units of observation. Let $Y_i$ denote the outcome (population/area smoke exposure per km$^2$) for observation $i$. Let $W_i$ denote the fire type, where $W_i=1$ represents a prescribed fire and $W_i=0$ represents a wildfire. Let $B_i$ denote the burn severity class ($B_i$: 1 = unburned-low, 2 = low, 3 = moderate, 4 = high). Let $\bm{X}_i$ be a vector consisting of the aforementioned topography, weather, and vegetation variables. Let $\bm{V}_i = (B_i, \bm{X}_i)$ be a vector of variables that may potentially moderate (that is, mitigate or worsen) the smoke exposure from prescribed fires or wildfires. Let $T_i$ indicate the occurrence/absence of fire (any fire type), where $T_i=1$ denotes a fire event, and $T_i=0$ denotes no fire. Adopting the potential outcomes framework in the causal inference literature \citep{holland1986statistics}, let $Y_i(T_i=t)$ denote the smoke exposure under event $T_i=t$. Because all of our observations are fire events, $Y_i(T_i=1) = Y_i$ is the observed smoke exposure of observation $i$, and $Y_i(T_i=0)$ is the \textit{counterfactual} smoke exposure had no fire occurred at the same area and time. We assume that the variables $Y(t)$ ($t=0, 1$) and $T$ are independent given the pre-fire covariates $\bm{X}$. We use uppercase letters to denote variables or vectors of variables (with boldface for vectors), and lowercase letters to denote their realizations.

We denote the expected smoke exposure from prescribed fires and wildfires as $S_1 = E_1[Y_i]$ and $S_0 = E_0[Y_i]$, respectively. The subscript of the expectation indicates the distribution under which the expectation is taken, e.g, $E_1[Y_i]$ is taken with respect to the distribution of smoke exposure among \emph{prescribed fires} (fire type $W_i=1$). Since we are interested in identifying factors in $\bm{V}$ that potentially modify the expected smoke exposure, we introduce the following notations: $\tau_{1}(\bm{v}) = E_1[Y_i| \bm{V}_i=\bm{v}]$, $\tau_{0}(\bm{v}) = E_0[Y_i|\bm{V}_i=\bm{v}]$. The conditioning on $\bm{V}_i=\bm{v}$ restricts the expectation to include only the sample with characteristic $\bm{v}$.

To address the challenges of dealing with the highly skewed outcome variable, we log-transformed the smoke exposure variable $Y_i$ in the statistical models: 
\begin{equation*}
    Y_i' = \log(Y_i+1).
\end{equation*}
For simplicity, in the following sections, we continue to use $Y_i$ (instead of $Y_i'$) to denote the log-transformed smoke exposure while noting that the original scale was used for comparisons  between prescribed fires and wildfires.

\subsection{Causal forest models}
We used causal forests to identify variables that underlie the heterogeneity in smoke exposure. Causal forests are an adaptation of random forests \citep{breiman2001random} to the problem of heterogeneous treatment effect estimation \citep{athey2019estimating,athey2019generalized,wager2018estimation}: instead of making predictions on an outcome, causal forests estimate the average causal effect given a set of covariates $\bm{V}$,
\begin{equation*}
    \tau_w(\bm{v}) = E_w[Y_i(T_i=1)-Y_i(T_i=0)| \bm{V}_i=\bm{v}], \quad w=0,1.
\end{equation*}
Two nuisance components, which could be separately estimated via any flexible predictive methods \citep{athey2019estimating}, are needed for building a causal forest:
\begin{itemize}
    \item Propensity score (the probability of fire given the pre-fire covariates $\bm{X}$): $e_w(\bm{x}) = E_w[T_i|\bm{X}_i=\bm{x}]$.
    \item Expected outcome (smoke exposure) given potential effect modifiers $\bm{V}$: $m_w(\bm{v}) = E_w[Y_i|\bm{V}_i=\bm{v}]$.
\end{itemize}
At a high level, the implementation of causal forests involves making out-of-bag predictions using the fitted propensity score and outcome models, and then incorporating these predictions to grow a causal forest. We provide a more technical review of causal forests in the Supplementary Information.

To account for potential correlations of outcomes (e.g., smoke exposure) within the same state and year, we applied cluster-robust statistical analyses, where a ``cluster" was defined as a state-year pair (e.g., CA 2018). Our prescribed fire sample consisted of 28 clusters, with sizes ranging from 2 to 216, while our wildfire sample included 36 clusters, with sizes ranging from 2 to 120. Details of all clusters and their sizes are provided in the Supplementary Information (Table B1). Therefore, the usual random forest sampling and prediction procedures needed to be modified to account for this clustered structure \citep{athey2019estimating}. Briefly, a subsample of clusters was drawn first, followed by drawing observations from each sampled cluster to build an individual tree. Additionally, an observation was considered out-of-bag if its cluster was not drawn. We refer to Athey and Wager \citep{athey2019estimating}, Sections 1.2 and 1.3, for more details on causal forests with clustered observations. 

We used the R package \texttt{grf} \citep{grf} to implement causal forests, using default tuning parameters and accounting for clustering by state-year. To estimate the average smoke exposure and cluster-robust standard error from prescribed fires ($S_1$) and identify moderating factors, we first duplicated the prescribed fire data. In this copy, we set the smoke exposure ($Y_i$) and event type ($T_i$) to zero, assuming no smoke exposure in the absence of fire ($Y_i(T_i=0) = 0$), while leaving all other variables unchanged. 
We then combined this ``counterfactual" copy with the original prescribed fire sample as the input for estimation. The \texttt{grf} package employs a variant of augmented inverse-propensity weighting \citep{robins1994estimation} to estimate $E_1[Y_i(T_i=1)-Y_i(T_i=0)]$. This quantity represented the expected smoke exposure $S_1 = E_1[Y_i]$, since $Y_i(T_i=0)$ was assumed to be zero. Estimation of the average smoke exposure from wildfires ($S_0$) followed the same procedure. The variable importance score for each variable in $\bm{V}$ was calculated as a weighted (by depth in the forest) sum of the number of times the variable was used for splitting when growing the causal forest. We note that due to the inherent randomness in sampling and variable selection, the importance scores and ranking of variables are not entirely fixed across implementations unless a random seed is set. However, the variables with the top five highest importance scores are generally consistent. Additionally, variable importance scores are not scale-invariant, i.e., variables identified as influential for smoke exposure on the log scale may differ from those identified on the original scale.

\subsection{Targeting operator characteristic}
Our study aimed to identify the conditions under which prescribed fires and wildfires resulted in higher or lower smoke exposure. To this end, we used the targeting operator characteristic (TOC; \citep{yadlowsky2025evaluating}) curve to assess the relationship between a single variable and smoke exposure per km$^2$.  

Let $V_{ij} \in \bm{V}_i$ be a continuous variable (indexed by $j$) in our covariate set. The TOC at $q$, evaluated on $V_{ij}$, is defined as (for $0 \leq q \leq 1$)
\begin{equation*}
    \text{TOC}_w(q) = E_w[Y_i(T_i=1)-Y_i(T_i=0)| F_w(V_{ij}) \leq q] - E_w[Y_i(T_i=1)-Y_i(T_i=0)],
\end{equation*}
where $F_w(\cdot)$ is the distribution function of $V_{ij}$ for fire type $w \in \{0,1\}$. The TOC compares the smoke exposure in the top fraction $q$ of observations with the smallest $V_{ij}$ values, to the overall smoke exposure. In other words, it evaluates the difference between targeting fires in areas with $V_{ij}$ values below a certain threshold versus distributing the fires randomly. We used the \texttt{grf} package to estimate the TOC along with cluster-robust standard errors. For prescribed fires ($w=1$) and wildfires ($w=0$), we plotted the estimated TOC curves (TOC($q$) plotted against $q$), focusing on variables with the top ten highest importance scores.

The Area Under the TOC (AUTOC) can be used to evaluate the presence of heterogeneity in smoke exposure across $V_{ij}$ \citep{yadlowsky2025evaluating}; it is defined as
\begin{equation*}
    \text{AUTOC} = \int_0^1 \text{TOC}(q)dq.
\end{equation*}
The AUTOC is expected to be non-zero if heterogeneity is present. The \texttt{grf} package provides estimates of the AUTOC and its standard error. We calculated the 95\% confidence interval for the AUTOC for each TOC plot -- an interval that includes zero indicates no significant evidence of heterogeneity in the smoke exposure across the evaluated variable. Essentially, the AUTOC assesses how good a variable is in ranking observations according to the estimated smoke exposure. 

\subsection{Comparison between prescribed fires and wildfires}
We compared the smoke exposure per km$^2$ between prescribed fires and wildfires while accounting for baseline environmental and geographical differences at the area and state levels. Specifically, we estimated the expected differences in smoke exposure between the counterfactual outcomes $Y_i(W_i=1)$ (smoke exposure if site $i$ had received a prescribed fire) and $Y_i(W_i=0)$ (smoke exposure if site $i$ had received a wildfire) for sites with a non-zero probability of experiencing either fire type (whereas in our previous analysis, we compared $Y_i(T_i=1)$ and $Y_i(T_i=0)$ within a single fire type). One challenge in comparing prescribed fires to wildfires is that they occur under markedly different environmental and geographical conditions; thus, a straightforward comparison of smoke exposure sample means may not accurately reflect the true differential impact. Moreover, most wildfires in our dataset originated in California, whereas the majority of prescribed fires were recorded in Florida. As a result, even after accounting for observed variations in topography, weather, and vegetation, unmeasured state-level factors may still confound the comparison. To mitigate these concerns, we treated the pre-fire covariates $\bm{X}$ and state indicators as confounders (denoted as $\bm{X^\prime}$) in assessing the effect of fire type (prescribed fire versus wildfire) on smoke exposure. Given the insufficient overlap in the distributions of these confounders between the two fire types, we estimated the overlap-weighted average causal effect \citep{li2018balancing}:
\begin{equation}\label{eq:WATE}
    \frac{E[e(\bm{X^\prime}) (1-e(\bm{X^\prime})) (Y(W_i=1) - Y(W_i=0))]}{E[e(\bm{X^\prime}) (1 - e(\bm{X^\prime})]},
\end{equation}
where $e(\bm{X^\prime}) = P[W_i = 1|\bm{X^\prime} = \bm{x^\prime}]$. A sample estimator of equation [\ref{eq:WATE}] is:
\begin{equation}\label{eq:WATE_est}
    \frac{\sum_i (1-e(\bm{X^\prime}_i))W_iY_i}{\sum_i (1-e(\bm{X^\prime}_i))W_i} - \frac{\sum_i e(\bm{X^\prime}_i)(1-W_i)Y_i}{\sum_i e(\bm{X^\prime}_i)(1-W_i)},
\end{equation}
Intuitively, a prescribed fire site ($W_i=1$) with a high probability of receiving a prescribed fire (large $e(\bm{X^\prime}_i)$) was assigned a small weight through the term $1-e(\bm{X^\prime}_i)$; a wildfire site ($W_i=0$) with a high probability of receiving a prescribed fire (large $e(\bm{X^\prime}_i)$) was assigned a large weight through $e(\bm{X^\prime}_i)$. This weighting scheme allowed us to target areas that have substantial probability of either fire type occurring \citep{li2018balancing}. We used the \texttt{grf} package to estimate the overlap-weighted average causal effect, and accounted for clustering by state-year to ensure valid standard error estimates.

\section*{Declarations}

\bmhead{Acknowledgments} 
Research reported in this publication was supported by NIH grants K01ES036202, RF1NS144268, P20AG093975, P30ES009089, P30ES023515, and UL1TR004419. The content is solely the responsibility of the authors and does not necessarily represent the official views of the NIH. XW is partially supported by the Calderone Award for Junior Faculty Development. MQ acknowledges the support from Minghua Zhang Faculty Career Catalyst Award from Stony Brook University. 

\bmhead{Author contributions}
T.C. participated in study design, methodology, data analyses, and manuscript drafting. M.Q. participated in conceptualization, data curation, and result interpretation. Y.W. participated in conceptualization and result interpretation. X.W. participated in securing resources, conceptualization, study design, methodology, and supervision. All authors contributed to manuscript review and editing.

\bmhead{Competing Interests}
X.W. has been employed at Meta since 2025. The work presented in this manuscript was unrelated to the work at Meta.

\bmhead{Data Availability}
All datasets utilized in this study are publicly available. The MTBS data were obtained from \url{https://www.mtbs.gov}; data on topography were obtained from \url{http://www.earthenv.org/topography}; data on vegetation land cover were obtained from \url{https://www.mrlc.gov/data}; weather data (precipitation, wind direction, wind velocity,
pressure, minimum and maximum temperature, minimum and
maximum relative humidity) were obtained from \url{https://www.drought.gov/data-maps-tools/gridded-surface-meteorological-gridmet-dataset}; The GlobFire data and fire smoke exposure were obtained from \url{https://github.com/jeffwen/smoke\_linking\_public}.

\bmhead{Computer code}
The code used for the analyses and the data shown in the figures are available at \url{https://github.com/tinghsuan-chang/fire\_moderation}.
\clearpage\renewcommand{\refname}{References}

\clearpage\appendix\section*{Supplementary Information}
\section{Technical Details}\label{secA1}
In this section, we expand on the methods presented in the main text by providing additional details on causal forests and the targeting operator characteristic. 

\subsection{Notation}
Let $i=1,...,n$ index the observations. Let $Y_i$ denote the smoke exposure (population smoke exposure per km$^2$ or area smoke exposure per km$^2$, as defined in the main text) for observation $i$. Let $\bm{X}_i$ be a vector of topography, weather, and vegetation variables observed prior to the fire event. Let $T_i$ indicate the occurrence/absence of fire, where $T_i=1$ denotes a fire event and $T_i=0$ denotes no fire. For simplicity, in this supplement, we exclude the fire type ($W_i$) and burn severity ($B_i$) variables to focus on explaining the methods. We also assume that the observations $\{Y_i, T_i, \bm{X}_i\}_{i=1,...,n}$ are independent and identically distributed\footnote{Smoke exposure is assumed to be arbitrarily correlated within the same state and year. For simplicity, we ignore this clustered structure here. However, we note that minor adaptations to the sampling and cross-fitting procedures -- essentially, treating ``clusters" as units -- are necessary to account for the clustered structure and ensure valid inference.}. Adopting the potential outcomes framework \citep{SI_holland1986statistics}, let $Y_i(t)$ denote the smoke exposure under event $T_i=t$; thus, $Y_i(1)$ is the smoke exposure had a fire occurred, and $Y_i(0)$ is the smoke exposure had no fire occurred. We use uppercase letters to denote variables or vectors of variables (with boldface for vectors), and lowercase letters to denote their realizations.

\subsection{Heterogeneous causal effects and causal forests}
The average causal effect (ACE) is defined as $\tau = E[Y_i(1)-Y_i(0)]$, representing the expected difference in smoke exposure between the occurrence of a fire and no fire. In our study sample, all observations are fire events, so each $Y_i(1) = Y_i$ is observed in the data, while $Y_i(0)$ is unobserved.

First, suppose that the causal effect is constant ($\tau'$) across subgroups (i.e., no heterogeneity in causal effects). Under this assumption, consider the following model:
\begin{equation}\label{eq: constant tau}
    Y_i = \tau' T_i + f(\bm{X}_i) + \varepsilon_i, \quad E[\varepsilon_i|\bm{X}_i, T_i]=0, 
\end{equation}
where $f(\cdot)$ is some unknown (possibly nonlinear) function of $\bm{X}_i$. Using Robinson's double residual method \citep{SI_robinson1988root}, we take the conditional expectation to both sides of [\ref{eq: constant tau}], which leads to
\begin{equation}\label{eq: constant tau cond exp}
    E[Y_i|\bm{X}_i=\bm{x}] = \tau' P[T_i=1|\bm{X}_i=\bm{x}] + f(\bm{x}). 
\end{equation}
Subtracting [\ref{eq: constant tau cond exp}] from [\ref{eq: constant tau}] yields 
\begin{equation}\label{eq: constant tau centered}
    Y_i - m(\bm{x}) = \tau' \times (T_i-e(\bm{x})) + \varepsilon_i, 
\end{equation}
where $m(\bm{x}) = E[Y_i|\bm{X}_i=\bm{x}]$ is the conditional mean outcome, and $e(\bm{x}) = P[T_i=1|\bm{X}_i=\bm{x}]$ is the propensity score\footnote{The propensity score model adjusts for potential confounders $\bm{X}$, which are covariates believed to influence both the occurrence of fire ($T$) and smoke exposure ($Y$). We make the conditional ignorability assumption $Y(t) \indep T \mid \bm{X}$ (for $t=0,1$), i.e., given the baseline covariates included in $\bm{X}$, the occurrence of fire should be independent of the potential outcomes. Thus, we may simply adjust for $\bm{X}$ in the propensity score model to eliminate confounding effects.}. Thus, the causal effect, $\tau'$, may be estimated by regressing the centered outcomes $Y_i-\hat{m}(\bm{x})$ on $T_i-\hat{e}(\bm{x})$ (residual-on-residual regression), where $\hat{m}(\bm{x})$ and $\hat{e}(\bm{x})$ are estimates of $m(\bm{x})$ and $e(\bm{x})$, respectively. van der Laan et al. \citep{SI_van2011cross} and Chernozhukov et al. \citep{SI_chernozhukov2018double} showed that if cross-fitting is done -- i.e., if we plug in the out-of-bag estimates $\hat{e}^{(-i)}(\bm{x})$ and $\hat{m}^{(-i)}(\bm{x})$ obtained by flexible machine learning methods (e.g., random forests), where the superscript $(-i)$ indicates that the $i$-th observation is excluded from the estimation -- it is possible to obtain root-n consistent estimates of $\tau'$. 

Now, suppose that the causal effect varies across subgroups. We can posit the following model:  
\begin{equation}\label{eq: non-constant tau}
    Y_i = \tau(\bm{X}_i) T_i + f(\bm{X}_i) + \varepsilon_i, \quad E[\varepsilon_i|\bm{X}_i, T_i]=0, 
\end{equation}
where $\tau(\bm{X}_i)$ is the \textit{conditional} average causal effect 
\begin{equation*}
    \tau(\bm{x}) = E[Y_i(1)-Y_i(0)|\bm{X}_i=\bm{x}].
\end{equation*}

Let $\mathcal{N}(\bm{x})$ denote a ``neighborhood" where the causal effect is a constant equal to the unknown quantity $\tau(\bm{x})$. The idea behind causal forests is to run a \textit{weighted} regression of $Y_i-\hat{m}^{(-i)}(\bm{x})$ on $T_i-\hat{e}^{(-i)}(\bm{x})$, with weights $\bm{1}\{\bm{X}_i \in \mathcal{N}(\bm{x})\}$, where $\bm{1}\{\cdot\}$ is the indicator function. In other words, for a target sample $\bm{X}_i=\bm{x}$, one could estimate $\tau(\bm{x})$ by performing the residual-on-residual regression on samples that share the same causal effect as the target. Below, we briefly describe how random forests are utilized to search for $\mathcal{N}(\bm{x})$ and how they are extended for constructing causal forests.

The goal of random forests is to predict the outcome based on a set of covariates \citep{SI_breiman2001random}, i.e., one may estimate $\mu(\bm{x}) = E[Y_i|\bm{X}_i=\bm{x}]$ (whereas causal forests estimate $\tau(\bm{x}) = E[Y_i(1)-Y_i(0)|\bm{X}_i=\bm{x}]$). The tree-building phase of random forests selects covariate splits that maximize the squared difference in outcome means between subgroups at each split, while in causal forests, the splits maximize the squared difference in estimated causal effects between subgroups. In a random forest, let $B$ be the number of trees, and for each tree $b = 1,...,B$, a subsample $\mathcal{S}_b \subseteq \{1,...,n\}$ of the observations is drawn. The estimate of $\mu(\bm{x})$ is obtained by averaging the predictions of the $B$ trees:
\begin{equation}\label{eq: rf}
    \begin{aligned}
        \hat{\mu}(\bm{x}) &= \frac{1}{B} \mathop{\sum}\limits_{b=1}^B \hat{\mu}^{(b)}(\bm{x}), \\ \hat{\mu}^{(b)}(\bm{x}) &= \mathop{\sum}\limits_{i=1}^n \frac{Y_i \textbf{1}\{\bm{X}_i \in L_b(\bm{x}), i \in \mathcal{S}_b\}}{|i: \bm{X}_i \in L_b(\bm{x}), i \in \mathcal{S}_b|},
    \end{aligned}
\end{equation}
where $L_b(\bm{x})$ denotes the leaf of the $b$-th tree containing $\bm{x}$. Following Athey and Wager \citep{SI_athey2019estimating}, [\ref{eq: rf}] can be re-expressed as a kernel form:
\begin{equation}\label{eq: rf kernal}
    \begin{aligned}
        \hat{\mu}(\bm{x}) &= \mathop{\sum}\limits_{i=1}^n \alpha_i(\bm{x})Y_i, \\
        \alpha_i(\bm{x}) &= \frac{1}{B} \mathop{\sum}\limits_{b=1}^B \frac{\textbf{1}\{\bm{X}_i \in L_b(\bm{x}), i \in \mathcal{S}_b\}}{|i: \bm{X}_i \in L_b(\bm{x}), i \in \mathcal{S}_b|}.
    \end{aligned}
\end{equation}
Here, $\alpha_i(\bm{x})$ measures how often $\bm{X}_i$ falls in the same ``neighborhood" as $\bm{x}$. Based on this expression, the $\alpha_i(\bm{x})$ from a causal forest can be used in the residual-on-residual regression to estimate $\tau(\bm{x})$. Essentially, causal forests run a weighted regression of $Y_i-\hat{m}^{(-i)}(\bm{x})$ on $T_i-\hat{e}^{(-i)}(\bm{x})$, using $\alpha_i(\bm{x})$ as weights.

To estimate the ACE, a doubly-robust approach, based on a variant of augmented inverse-propensity weighting (AIPW; \citep{SI_robins1994estimation}), can be employed:
\begin{equation}\label{eq: AIPW}
    \hat{\tau} = \frac{1}{n}\sum_{i=1}^n \bigg(\hat{\tau}^{(-i)}(\bm{X}_i) + \frac{T_i-\hat{e}^{(-i)}(\bm{X}_i)}{\hat{e}^{(-i)}(\bm{X}_i)(1-\hat{e}^{(-i)}(\bm{X}_i))} \Big(Y_i-\hat{m}^{(-i)}(\bm{X}_i)-\big(T_i-\hat{e}^{(-i)}(\bm{X}_i)\big)\hat{\tau}^{(-i)}(\bm{X}_i)\Big)\bigg).
\end{equation}
Under some regularity conditions, this AIPW estimator of $\tau$ is shown to be asymptotically efficient (i.e., has the lowest possible asymptotic variance) if the nuisance components ($e(\bm{X})$ and $m(\bm{X})$) are non-parametrically estimated \citep{SI_van2011cross,SI_chernozhukov2018double}.

\subsection{Assessing heterogeneity}
Suppose $T_i=1$ indicates the administration of a prescribed fire ($T_i=0$ indicates no fire;  prescribed fire is used here as an example for illustration). Our goal is to identify which sites should be prioritized for prescribed fire administration using a ``prioritization rule", $S(\bm{X}_i)$, which is a function that maps site characteristics to a scalar value or ``score". Ideally, we want $S(\bm{X}_i)$ to assign a high score to sites expected to produce less smoke (specifically, small ACE) under a prescribed fire and a low score to sites likely to produce excessive smoke (large ACE). In our study, we are interested in individual covariates (indexed by $j$), $X_{ij} \in \bm{X}_i$; specifically, we aim to assess whether a simple prioritization rule $S(\bm{X}_i)=X_{ij}$ (or $-X_{ij}$ if smoke exposure is believed to increase with $X_{ij}$) is effective in identifying sites likely to produce lower smoke impacts from prescribed fires and should therefore be prioritized. Below we briefly review the rank-weighted average treatment effect (RATE) metrics introduced by Yadlowsky and colleagues \citep{SI_yadlowsky2025evaluating}, which can be used to assess the performance of a prioritization rule $S(\bm{X}_i)$. We begin by defining the Targeting Operator Characteristic (TOC) curve and its interpretation within the context of our study.

\subsubsection{Targeting Operator Characteristic}
Let $F_{S(\bm{X})}$ be the  distribution function of prioritization rule $S(\bm{X}_i)$. For $q \in [0,1]$, we define the TOC at quantile $q$ as \citep{SI_yadlowsky2025evaluating}
\begin{equation}\label{eq: TOC}
    \text{TOC}(q) = E[Y_i(1)-Y_i(0)|S(\bm{X}_i) \leq F^{-1}_{S(\bm{X})}(q)] - E[Y_i(1)-Y_i(0)].
\end{equation}
The first term on the right-hand side is the ACE for the $q$\% of sites with the lowest prioritization scores, and the second term is the overall ACE. If $S(\bm{X}_i)=X_{ij}$, 
\begin{equation}\label{eq: TOC simple}
    \text{TOC}(q) = E[Y_i(1)-Y_i(0)|X_{ij} \leq F^{-1}_{S(\bm{X})}(q)] - E[Y_i(1)-Y_i(0)],
\end{equation}
TOC($q$) compares the ACE in the subgroup with $X_{ij}$ in the lowest $q$\% to the overall ACE; TOC($q$) is expected to be negative if the ACE of prescribed fires is believed to (monotonically) decrease as $X_{ij}$ increases. The TOC curve plots TOC($q$) against all $q \in [0,1]$. When the prioritization rule $S(\bm{X}_{i})$ simply outputs the value of a single covariate $X_{ij}$, the TOC curve provides a visual tool to examine heterogeneity in the ACE across quantiles of $X_{ij}$. 

\subsubsection{RATE metrics}
Rank-weighted Average Treatment Effects (RATE) metrics allow us to quantify and evaluate the performance of a prioritization rule \citep{SI_yadlowsky2025evaluating}. In our simple case, where $S(\bm{X}_{i})=X_{ij}$, the RATE metrics indicate how effectively  rankings based on a specific covariate $X_{ij}$ can be used to prioritize sites for prescribed fires. RATE metrics are defined as weighted averages of the TOC: for a given weight function $\alpha: (0,1] \to \mathbb{R}$, the RATE of prioritization rule $S$ is
\begin{equation}\label{eq: RATE}
    \theta_\alpha(S) = \int^1_0 \alpha(q)\text{TOC}(q;S)dq.
\end{equation}
In our study, we use the constant weight function $\alpha(q)=1$ and thus 
\begin{equation}\label{eq: AUTOC}
    \theta_\alpha(S) = \int^1_0 \text{TOC}(q;S)dq,
\end{equation}
which is the Area Under the TOC (AUTOC). Another commonly used weight function is $\alpha(q)=q$, which gives rise to the Qini coefficient \citep{SI_radcliffe2007using}. Intuitively, if there is little heterogeneity in the ACE of prescribed fires across quantiles of the covariate $X_{ij}$, the AUTOC of prioritization rule $S(\bm{X}_i)=X_{ij}$ will be small, reaching zero in the case of no heterogeneity. Thus, to examine the presence of effect heterogeneity across $X_{ij}$ (i.e., effect moderation by $X_{ij}$), we can estimate its AUTOC and test it against the sharp null of AUTOC $=0$. For the technical derivations and properties of general RATE estimators, we refer readers to Yadlowsky et al. \citep{SI_yadlowsky2025evaluating}. The doubly-robust estimator of RATE metrics is asymptotically normal under regularity conditions, i.e., $\sqrt{n}(\hat{\theta}-\theta) \to_d N(0,\sigma_\theta)$, where the asymptotic standard error $\sigma_\theta$ can be approximated using half-sample bootstrap \citep{SI_yadlowsky2025evaluating}.

\section{Supplemental Results}
\begin{figure}[H]
\centering
\includegraphics[width=16cm,height=6cm]{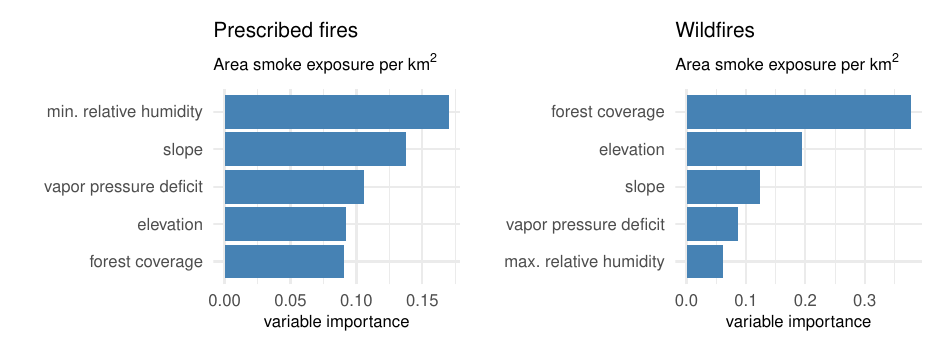}
\caption{Variable importance scores in moderating the area smoke exposure (log \mugm) per km$^2$ of prescribed fires (left) and wildfires (right).}\label{fig:varimp_area}
\end{figure}

\begin{figure}[H]
\centering
\includegraphics[width=\textwidth,height=.75\textheight,keepaspectratio]{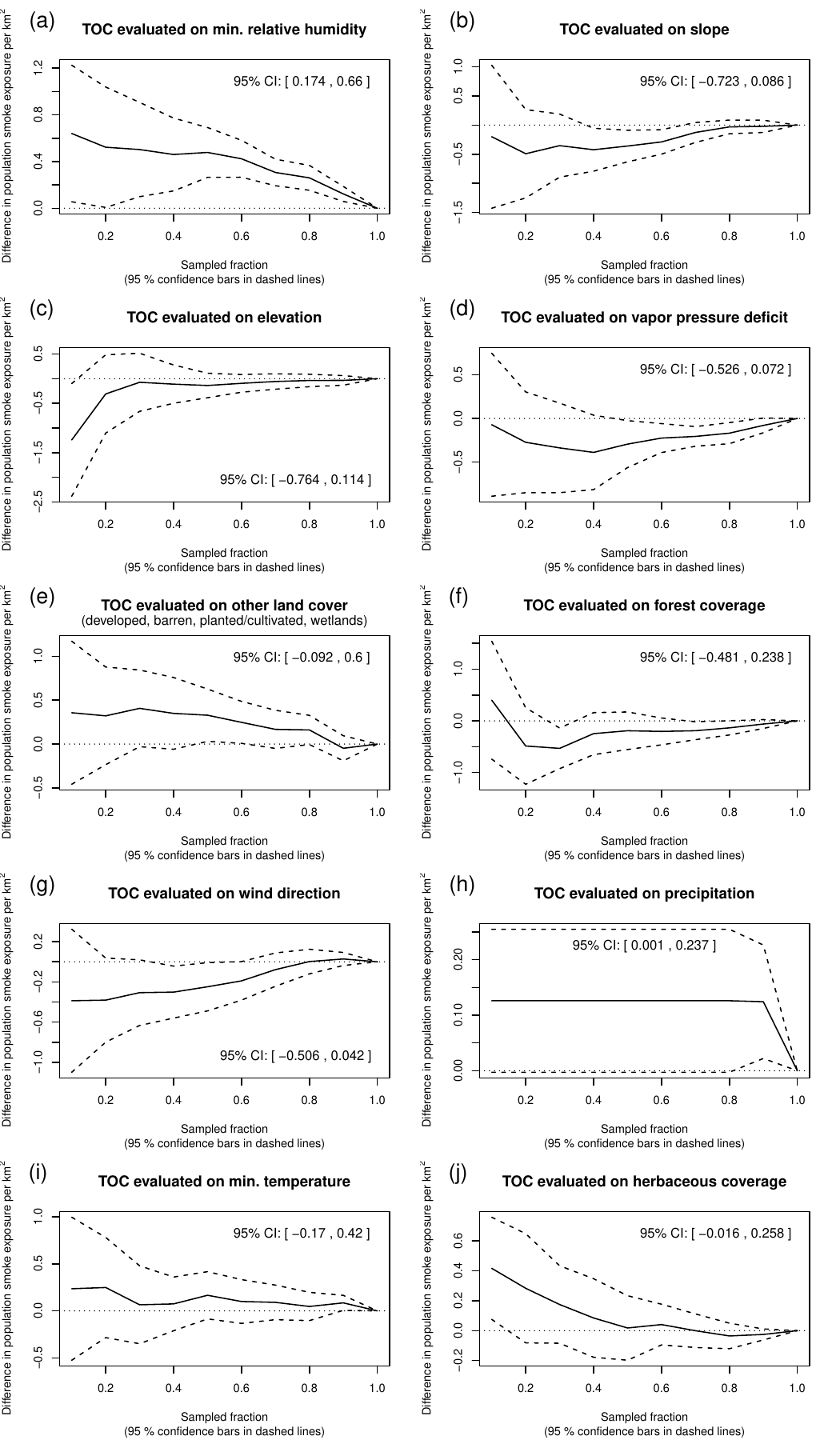}
\caption{Targeting Operator Characteristic (TOC) curves for the variation in population smoke exposure (log person-\mugm) per km$^2$ from prescribed fires attributed to each variable. The y-axis shows how smoke exposure varies in regions with different values of a specific variable  compared to the overall average. A downward-trending curve indicates that lower values of the evaluated variable are associated with higher smoke exposure. The 95\% confidence interval (CI) of the area under the TOC curve (AUTOC) is displayed in each plot.}\label{fig:TOC pres pop}
\end{figure}

\begin{figure}[H]
\centering
\includegraphics[width=\textwidth,height=.75\textheight,keepaspectratio]{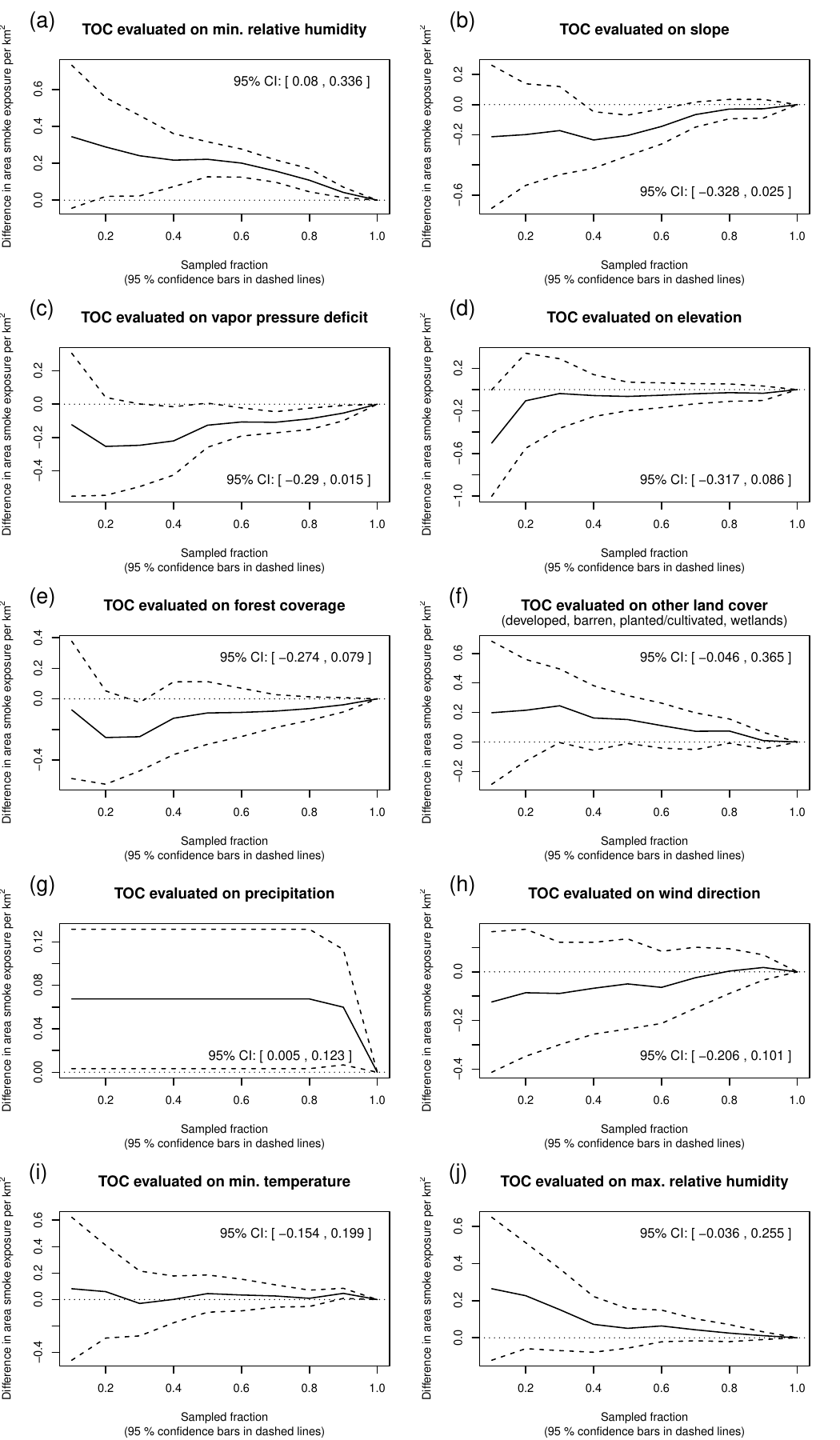}
\caption{Targeting Operator Characteristic (TOC) curves for the variation in area smoke exposure (log \mugm) per km$^2$ from prescribed fires attributed to each variable. The y-axis shows how smoke exposure varies in regions with different values of a specific variable (e.g., minimal relative humidity and precipitation) compared to the overall average. A downward-trending curve indicates that lower values of the evaluated variable are associated with higher smoke exposure. The 95\% confidence interval (CI) of the area under the TOC curve (AUTOC) is displayed in each plot.}\label{fig:TOC pres area}
\end{figure}

\begin{figure}[H]
\centering
\includegraphics[width=\textwidth,height=.75\textheight,keepaspectratio]{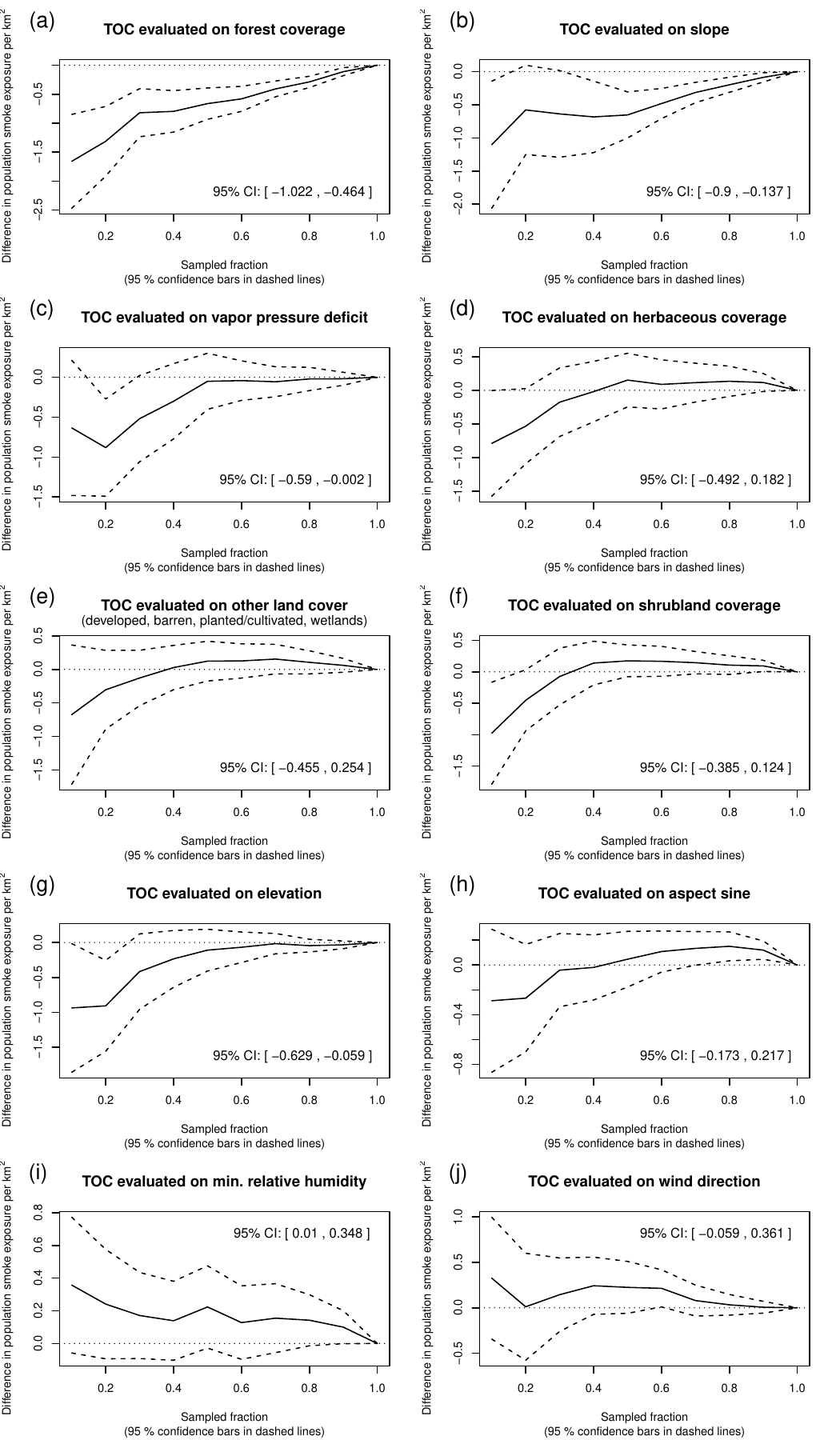}
\caption{Targeting Operator Characteristic (TOC) curves for the variation in population smoke exposure (log person-\mugm) per km$^2$ from wildfires attributed to each variable. The y-axis shows how smoke exposure varies in regions with different values of a specific variable  compared to the overall average. A downward-trending curve indicates that lower values of the evaluated variable are associated with higher smoke exposure. The 95\% confidence interval (CI) of the area under the TOC curve (AUTOC) is displayed in each plot.}\label{fig:TOC wild pop}
\end{figure}

\begin{figure}[H]
\centering
\includegraphics[width=\textwidth,height=.75\textheight,keepaspectratio]{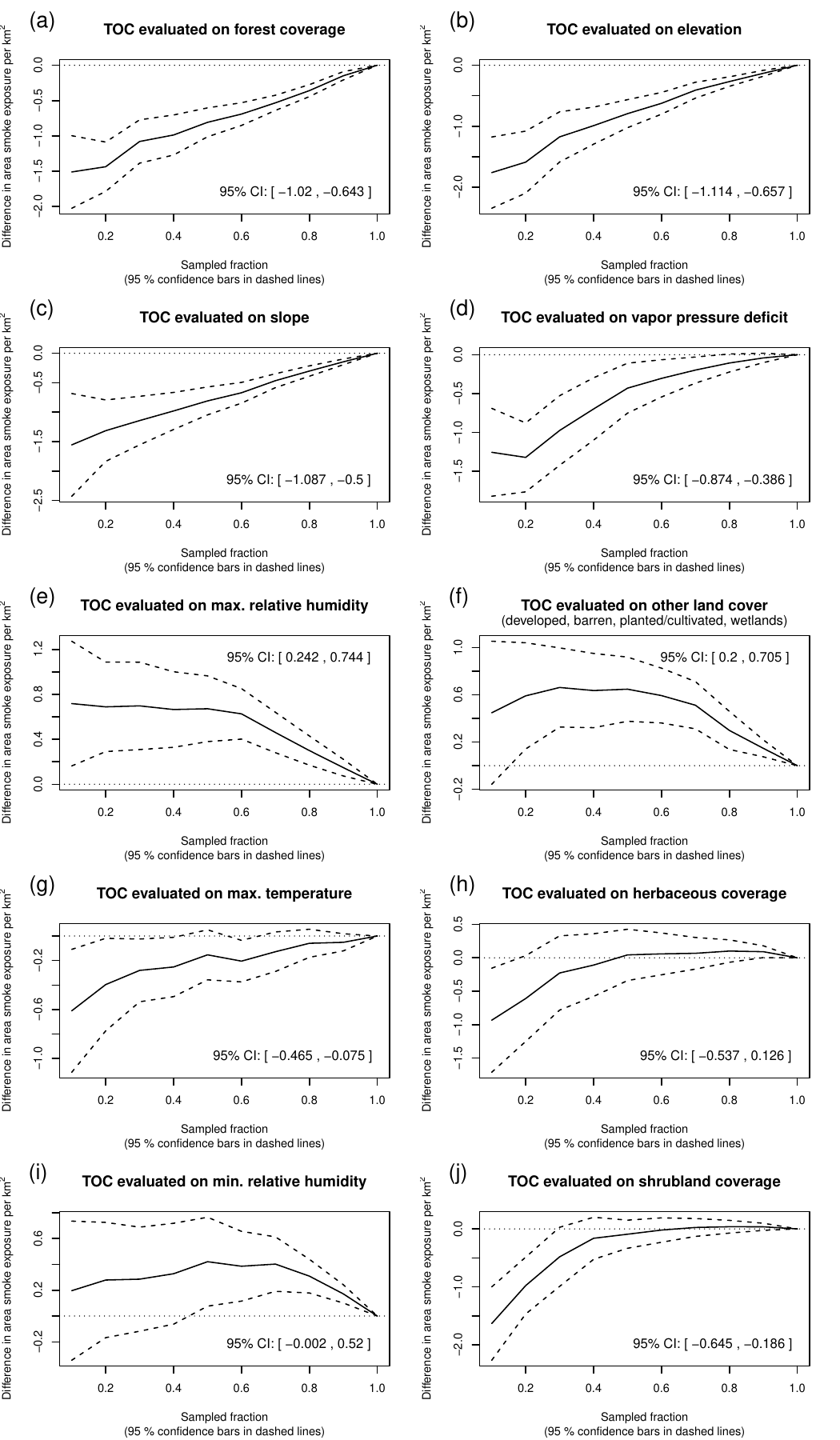}
\caption{Targeting Operator Characteristic (TOC) curves for the variation in area smoke exposure (log \mugm) per km$^2$ from wildfires attributed to each variable. The y-axis shows how smoke exposure varies in regions with different values of a specific variable  compared to the overall average. A downward-trending curve indicates that lower values of the evaluated variable are associated with higher smoke exposure. The 95\% confidence interval (CI) of the area under the TOC curve (AUTOC) is displayed in each plot.}\label{fig:TOC wild area}
\end{figure}

Table~\ref{tab:clusters} shows the number of fire events in each state (California, Florida, Georgia) and year (2006 to 2020) in our study sample. Figure~\ref{fig:varimp_area} shows the top five variables that contributed the most to the overall heterogeneity in area smoke exposure ($\log$ \mugm) per km$^2$ from prescribed fires and wildfires, along with their variable importance scores. Key variables include minimum relative humidity, slope, and vapor pressure deficit for prescribed fires, as well as forest coverage, elevation, and slope for wildfires. In Figures~\ref{fig:TOC pres pop}-\ref{fig:TOC wild area}, we show the targeting operator characteristic (TOC) curves that assess the direction of moderating effect of each variable (those among the ten highest-ranked in variable importance scores) on smoke exposure per km$^2$. For prescribed fires, area smoke exposure ($\log$ \mugm) tended to be lower at sites with higher minimum relative humidity and at sites experiencing very high levels of
precipitation, aligning with findings for population smoke exposure ($\log$ person-\mugm). For wildfires, the findings for area smoke exposure ($\log$ \mugm) were also largely consistent with those for population smoke exposure ($\log$ person-\mugm) -- lower area smoke exposure ($\log$ \mugm) was linked to sites with less forest coverage, lower elevations, gentler slopes, lower vapor pressure deficits, higher maximum relative humidity, lower maximum temperature, and less shrubland coverage.

\newpage

\begin{table}[hbt!]
\centering
\caption{Number of fire events in each state-year in our study sample}
\begin{tabular}{rrr}
  \hline
 & Prescribed fire & Wildfire \\ 
  \hline
  CA2006 &   2 &  68 \\ 
  CA2007 &   0 &  68 \\ 
  CA2008 &   2 & 120 \\ 
  CA2009 &   2 &  60 \\ 
  CA2010 &   0 &  30 \\ 
  CA2011 &   0 &  24 \\ 
  CA2012 &   0 &  58 \\ 
  CA2013 &   0 &  54 \\ 
  CA2014 &   0 &  58 \\ 
  CA2015 &   0 &  70 \\ 
  CA2016 &   0 &  72 \\ 
  CA2017 &   2 & 116 \\ 
  CA2018 &   0 &  48 \\ 
  CA2019 &   0 &  24 \\ 
  CA2020 &   0 &  92 \\ 
  FL2006 &   4 &  18 \\ 
  FL2007 &  44 &  38 \\ 
  FL2008 & 144 &  20 \\ 
  FL2009 & 172 &  10 \\ 
  FL2010 & 172 &   8 \\ 
  FL2011 & 216 &  72 \\ 
  FL2012 & 114 &  12 \\ 
  FL2013 &  32 &  16 \\ 
  FL2014 &  28 &   4 \\ 
  FL2015 &  24 &   6 \\ 
  FL2016 &  36 &   0 \\ 
  FL2017 &  28 &  52 \\ 
  FL2018 &  42 &  16 \\ 
  FL2019 &  26 &   4 \\ 
  FL2020 &  20 &   8 \\ 
  GA2006 &   0 &   0 \\ 
  GA2007 &   0 &  20 \\ 
  GA2008 &   6 &   4 \\ 
  GA2009 &   0 &   0 \\ 
  GA2010 &   6 &   2 \\ 
  GA2011 &   2 &  20 \\ 
  GA2012 &   2 &   0 \\ 
  GA2013 &   0 &   0 \\ 
  GA2014 &   8 &   2 \\ 
  GA2015 &   4 &   0 \\ 
  GA2016 &   6 &   4 \\ 
  GA2017 &   4 &   2 \\ 
  GA2018 &  10 &   0 \\ 
  GA2019 &   0 &   0 \\ 
   \hline
\end{tabular}
\label{tab:clusters}
\end{table}

\newpage

\newpage

\newpage

\clearpage\renewcommand{\refname}{Supplementary References}

\begin{thebibliography}{45}
\providecommand{\natexlab}[1]{#1}
\providecommand{\url}[1]{\texttt{#1}}
\expandafter\ifx\csname urlstyle\endcsname\relax
  \providecommand{\doi}[1]{doi: #1}\else
  \providecommand{\doi}{doi: \begingroup \urlstyle{rm}\Url}\fi

\bibitem[RN2(2024)]{RN21}
State of the air.
\newblock \emph{American Lung Association}, 2024.

\bibitem[Abatzoglou(2013)]{abatzoglou2013development}
John~T Abatzoglou.
\newblock Development of gridded surface meteorological data for ecological
  applications and modelling.
\newblock \emph{International journal of climatology}, 33\penalty0
  (1):\penalty0 121--131, 2013.

\bibitem[Amatulli et~al.(2018)Amatulli, Domisch, Tuanmu, Parmentier, Ranipeta,
  Malczyk, and Jetz]{amatulli2018suite}
Giuseppe Amatulli, Sami Domisch, Mao-Ning Tuanmu, Benoit Parmentier, Ajay
  Ranipeta, Jeremy Malczyk, and Walter Jetz.
\newblock A suite of global, cross-scale topographic variables for
  environmental and biodiversity modeling.
\newblock \emph{Scientific data}, 5\penalty0 (1):\penalty0 1--15, 2018.

\bibitem[Art{\'e}s et~al.(2019)Art{\'e}s, Oom, De~Rigo, Durrant, Maianti,
  Libert{\`a}, and San-Miguel-Ayanz]{artes2019global}
Tom{\`a}s Art{\'e}s, Duarte Oom, Daniele De~Rigo, Tracy~Houston Durrant,
  Pieralberto Maianti, Giorgio Libert{\`a}, and Jes{\'u}s San-Miguel-Ayanz.
\newblock A global wildfire dataset for the analysis of fire regimes and fire
  behaviour.
\newblock \emph{Scientific data}, 6\penalty0 (1):\penalty0 296, 2019.

\bibitem[Athey and Wager(2019{\natexlab{a}})]{RN37}
Susan Athey and Stefan Wager.
\newblock Estimating treatment effects with causal forests: An application.
\newblock \emph{Observational studies}, 5\penalty0 (2):\penalty0 37--51,
  2019{\natexlab{a}}.
\newblock ISSN 2767-3324.

\bibitem[Athey and Wager(2019{\natexlab{b}})]{athey2019estimating}
Susan Athey and Stefan Wager.
\newblock Estimating treatment effects with causal forests: An application.
\newblock \emph{Observational studies}, 5\penalty0 (2):\penalty0 37--51,
  2019{\natexlab{b}}.

\bibitem[Athey et~al.(2019)Athey, Tibshirani, and Wager]{athey2019generalized}
Susan Athey, Julie Tibshirani, and Stefan Wager.
\newblock {Generalized random forests}.
\newblock \emph{The Annals of Statistics}, 47\penalty0 (2):\penalty0 1148 --
  1178, 2019.

\bibitem[Boisramé et~al.(2017)Boisramé, Thompson, Collins, and
  Stephens]{RN29}
Gabrielle Boisramé, Sally Thompson, Brandon Collins, and Scott Stephens.
\newblock Managed wildfire effects on forest resilience and water in the sierra
  nevada.
\newblock \emph{Ecosystems}, 20:\penalty0 717--732, 2017.
\newblock ISSN 1432-9840.

\bibitem[Breiman(2001)]{breiman2001random}
Leo Breiman.
\newblock Random forests.
\newblock \emph{Machine learning}, 45:\penalty0 5--32, 2001.

\bibitem[Brenner and Wade()]{RN36}
Jim Brenner and Dale Wade.
\newblock Florida’s revised prescribed fire law: Protection for responsible
  burners.
\newblock In \emph{Proceedings of Fire Conference}, pages 132--136.

\bibitem[Brown et~al.(2023)Brown, Hanley, Mahesh, Reed, Strenfel, Davis,
  Kochanski, and Clements]{RN26}
Patrick~T Brown, Holt Hanley, Ankur Mahesh, Colorado Reed, Scott~J Strenfel,
  Steven~J Davis, Adam~K Kochanski, and Craig~B Clements.
\newblock Climate warming increases extreme daily wildfire growth risk in
  california.
\newblock \emph{Nature}, 621\penalty0 (7980):\penalty0 760--766, 2023.
\newblock ISSN 0028-0836.

\bibitem[Burke et~al.(2023)Burke, Childs, de~la Cuesta, Qiu, Li, Gould,
  Heft-Neal, and Wara]{burke2023contribution}
Marshall Burke, Marissa~L Childs, Brandon de~la Cuesta, Minghao Qiu, Jessica
  Li, Carlos~F Gould, Sam Heft-Neal, and Michael Wara.
\newblock The contribution of wildfire to pm2. 5 trends in the usa.
\newblock \emph{Nature}, 622\penalty0 (7984):\penalty0 761--766, 2023.

\bibitem[Childs et~al.(2022)Childs, Li, Wen, Heft-Neal, Driscoll, Wang, Gould,
  Qiu, Burney, and Burke]{childs2022daily}
Marissa~L Childs, Jessica Li, Jeffrey Wen, Sam Heft-Neal, Anne Driscoll,
  Sherrie Wang, Carlos~F Gould, Minghao Qiu, Jennifer Burney, and Marshall
  Burke.
\newblock Daily local-level estimates of ambient wildfire smoke pm2. 5 for the
  contiguous us.
\newblock \emph{Environmental Science \& Technology}, 56\penalty0
  (19):\penalty0 13607--13621, 2022.

\bibitem[Eidenshink et~al.(2007)Eidenshink, Schwind, Brewer, Zhu, Quayle, and
  Howard]{eidenshink2007project}
Jeff Eidenshink, Brian Schwind, Ken Brewer, Zhi-Liang Zhu, Brad Quayle, and
  Stephen Howard.
\newblock A project for monitoring trends in burn severity.
\newblock \emph{Fire ecology}, 3:\penalty0 3--21, 2007.

\bibitem[Engel(2013)]{engel2013perverse}
Kirsten~H Engel.
\newblock Perverse incentives: The case of wildfire smoke regulation.
\newblock \emph{Ecology LQ}, 40:\penalty0 623, 2013.

\bibitem[Fann et~al.(2018)Fann, Alman, Broome, Morgan, Johnston, Pouliot, and
  Rappold]{RN22}
Neal Fann, Breanna Alman, Richard~A Broome, Geoffrey~G Morgan, Fay~H Johnston,
  George Pouliot, and Ana~G Rappold.
\newblock The health impacts and economic value of wildland fire episodes in
  the us: 2008–2012.
\newblock \emph{Science of the Total Environment}, 610:\penalty0 802--809,
  2018.
\newblock ISSN 0048-9697.

\bibitem[{Forest Management Task Force}(2021)]{forest2021california}
{Forest Management Task Force}.
\newblock California’s wildfire and forest resilience action plan, 2021.

\bibitem[Gould et~al.(2023)Gould, Heft-Neal, Prunicki, Aguilera, Burke, and
  Nadeau]{RN4}
Carlos~F Gould, Sam Heft-Neal, Mary Prunicki, Juan Aguilera, Marshall Burke,
  and Kari Nadeau.
\newblock Health effects of wildfire smoke exposure.
\newblock \emph{Annual Review of Medicine}, 75, 2023.
\newblock ISSN 0066-4219.

\bibitem[Holland(1986)]{holland1986statistics}
Paul~W Holland.
\newblock Statistics and causal inference.
\newblock \emph{Journal of the American statistical Association}, 81\penalty0
  (396):\penalty0 945--960, 1986.

\bibitem[Huang et~al.(2018)Huang, Zhang, Chan, Kondragunta, Russell, and
  Odman]{huang2018burned}
Ran Huang, Xiaoyang Zhang, Daniel Chan, Shobha Kondragunta, Armistead~G
  Russell, and M~Talat Odman.
\newblock Burned area comparisons between prescribed burning permits in
  southeastern united states and two satellite-derived products.
\newblock \emph{Journal of Geophysical Research: Atmospheres}, 123\penalty0
  (9):\penalty0 4746--4757, 2018.

\bibitem[Imbens and Rubin(2015)]{imbens2015causal}
Guido~W Imbens and Donald~B Rubin.
\newblock \emph{Causal inference in statistics, social, and biomedical
  sciences}.
\newblock Cambridge university press, Cambridge, United Kingdom, 2015.

\bibitem[Janssen et~al.(2023)Janssen, Jones, Finney, van~der Werf, van Wees,
  Xu, and Veraverbeke]{RN28}
Thomas~AJ Janssen, Matthew~W Jones, Declan Finney, Guido~R van~der Werf, Dave
  van Wees, Wenxuan Xu, and Sander Veraverbeke.
\newblock Extratropical forests increasingly at risk due to lightning fires.
\newblock \emph{Nature Geoscience}, 16\penalty0 (12):\penalty0 1136--1144,
  2023.
\newblock ISSN 1752-0894.

\bibitem[Kelp et~al.(2024)Kelp, Burke, Qiu, Higuera-Mendieta, Liu, and
  Diffenbaugh]{kelp2024efficacy}
Makoto Kelp, Marshall Burke, Minghao Qiu, Ivan Higuera-Mendieta, Tianjia Liu,
  and Noah~S Diffenbaugh.
\newblock Efficacy of recent prescribed burning and land management on wildfire
  burn severity and smoke emissions in the western united states.
\newblock \emph{EarthArXiv}, 2024.

\bibitem[Li et~al.(2018)Li, Morgan, and Zaslavsky]{li2018balancing}
Fan Li, Kari~Lock Morgan, and Alan~M Zaslavsky.
\newblock Balancing covariates via propensity score weighting.
\newblock \emph{Journal of the American Statistical Association}, 113\penalty0
  (521):\penalty0 390--400, 2018.

\bibitem[Littell et~al.(2016)Littell, Peterson, Riley, Liu, and Luce]{RN27}
Jeremy~S Littell, David~L Peterson, Karin~L Riley, Yongquiang Liu, and
  Charles~H Luce.
\newblock A review of the relationships between drought and forest fire in the
  united states.
\newblock \emph{Global Change Biology}, 22\penalty0 (7):\penalty0 2353--2369,
  2016.
\newblock ISSN 1354-1013.

\bibitem[Liu et~al.(2016)Liu, Mickley, Sulprizio, Dominici, Yue, Ebisu,
  Anderson, Khan, Bravo, and Bell]{RN18}
Jia~Coco Liu, Loretta~J Mickley, Melissa~P Sulprizio, Francesca Dominici,
  Xu~Yue, Keita Ebisu, Georgiana~Brooke Anderson, Rafi~FA Khan, Mercedes~A
  Bravo, and Michelle~L Bell.
\newblock Particulate air pollution from wildfires in the western us under
  climate change.
\newblock \emph{Climatic Change}, 138:\penalty0 655--666, 2016.
\newblock ISSN 0165-0009.

\bibitem[Moritz et~al.(2014)Moritz, Batllori, Bradstock, Gill, Handmer,
  Hessburg, Leonard, McCaffrey, Odion, Schoennagel, et~al.]{moritz2014learning}
Max~A Moritz, Enric Batllori, Ross~A Bradstock, A~Malcolm Gill, John Handmer,
  Paul~F Hessburg, Justin Leonard, Sarah McCaffrey, Dennis~C Odion, Tania
  Schoennagel, et~al.
\newblock Learning to coexist with wildfire.
\newblock \emph{Nature}, 515\penalty0 (7525):\penalty0 58--66, 2014.

\bibitem[Pacheco and Claro(2021)]{RN34}
Renata~Martins Pacheco and João Claro.
\newblock Prescribed burning as a cost-effective way to address climate change
  and forest management in mediterranean countries.
\newblock \emph{Annals of Forest Science}, 78\penalty0 (4):\penalty0 1--11,
  2021.
\newblock ISSN 1297-966X.

\bibitem[Parks et~al.(2014)Parks, Miller, Nelson, and
  Holden]{parks2014previous}
Sean~A Parks, Carol Miller, Cara~R Nelson, and Zachary~A Holden.
\newblock Previous fires moderate burn severity of subsequent wildland fires in
  two large western us wilderness areas.
\newblock \emph{Ecosystems}, 17:\penalty0 29--42, 2014.

\bibitem[Qiu et~al.(2024)Qiu, Li, Gould, Jing, Kelp, Childs, Wen, Xie, Lin,
  Kiang, et~al.]{qiu2024wildfire}
Minghao Qiu, Jessica Li, Carlos Gould, Renzhi Jing, Makoto Kelp, Marissa
  Childs, Jeff Wen, Yuanyu Xie, Meiyun Lin, Mathew Kiang, et~al.
\newblock Wildfire smoke exposure and mortality burden in the us under future
  climate change.
\newblock \emph{EarthArXiv}, 2024.

\bibitem[Reid et~al.(2016)Reid, Brauer, Johnston, Jerrett, Balmes, and
  Elliott]{RN30}
Colleen~E Reid, Michael Brauer, Fay~H Johnston, Michael Jerrett, John~R Balmes,
  and Catherine~T Elliott.
\newblock Critical review of health impacts of wildfire smoke exposure.
\newblock \emph{Environmental Health Perspectives}, 124\penalty0 (9):\penalty0
  1334--1343, 2016.
\newblock ISSN 0091-6765.

\bibitem[Reisen et~al.(2015)Reisen, Duran, Flannigan, Elliott, and
  Rideout]{RN31}
Fabienne Reisen, Sandra~M Duran, Mike Flannigan, Catherine Elliott, and Karen
  Rideout.
\newblock Wildfire smoke and public health risk.
\newblock \emph{International Journal of Wildland Fire}, 24\penalty0
  (8):\penalty0 1029--1044, 2015.
\newblock ISSN 1448-5516.

\bibitem[Riddle(2023)]{RN19}
Anne~A. Riddle.
\newblock \emph{Wildfire statistics}.
\newblock Congressional Research Service, Washington, DC, 2023.

\bibitem[Robins et~al.(1994)Robins, Rotnitzky, and Zhao]{robins1994estimation}
James~M Robins, Andrea Rotnitzky, and Lue~Ping Zhao.
\newblock Estimation of regression coefficients when some regressors are not
  always observed.
\newblock \emph{Journal of the American statistical Association}, 89\penalty0
  (427):\penalty0 846--866, 1994.

\bibitem[Service(2022)]{usda2022confronting}
USDA~Forest Service.
\newblock “confronting the wildfire crisis: A strategy for protecting
  communities and improving resilience in america’s forests, 2022.

\bibitem[Taylor et~al.(2014)Taylor, McCarthy, and
  Lindenmayer]{taylor2014nonlinear}
Chris Taylor, Michael~A McCarthy, and David~B Lindenmayer.
\newblock Nonlinear effects of stand age on fire severity.
\newblock \emph{Conservation Letters}, 7\penalty0 (4):\penalty0 355--370, 2014.

\bibitem[Thompson et~al.(2007)Thompson, Spies, and Ganio]{thompson2007reburn}
Jonathan~R Thompson, Thomas~A Spies, and Lisa~M Ganio.
\newblock Reburn severity in managed and unmanaged vegetation in a large
  wildfire.
\newblock \emph{Proceedings of the National Academy of Sciences}, 104\penalty0
  (25):\penalty0 10743--10748, 2007.

\bibitem[Tibshirani et~al.(2024)Tibshirani, Athey, Sverdrup, and Wager]{grf}
Julie Tibshirani, Susan Athey, Erik Sverdrup, and Stefan Wager.
\newblock \emph{grf: Generalized Random Forests}, 2024.
\newblock URL \url{https://github.com/grf-labs/grf}.
\newblock R package version 2.3.2, commit
  f79679d72124db505b92aca0be0ab2a694b70749.

\bibitem[Turner et~al.(2003)Turner, Romme, and Tinker]{turner2003surprises}
Monica~G Turner, William~H Romme, and Daniel~B Tinker.
\newblock Surprises and lessons from the 1988 yellowstone fires.
\newblock \emph{Frontiers in Ecology and the Environment}, 1\penalty0
  (7):\penalty0 351--358, 2003.

\bibitem[Wager and Athey(2018)]{wager2018estimation}
Stefan Wager and Susan Athey.
\newblock Estimation and inference of heterogeneous treatment effects using
  random forests.
\newblock \emph{Journal of the American Statistical Association}, 113\penalty0
  (523):\penalty0 1228--1242, 2018.

\bibitem[Wen et~al.(2023)Wen, Heft-Neal, Baylis, Boomhower, and
  Burke]{wen2023quantifying}
Jeff Wen, Sam Heft-Neal, Patrick Baylis, Judson Boomhower, and Marshall Burke.
\newblock Quantifying fire-specific smoke exposure and health impacts.
\newblock \emph{Proceedings of the National Academy of Sciences}, 120\penalty0
  (51):\penalty0 e2309325120, 2023.

\bibitem[Wettstein et~al.(2018)Wettstein, Hoshiko, Fahimi, Harrison, Cascio,
  and Rappold]{RN32}
Zachary~S Wettstein, Sumi Hoshiko, Jahan Fahimi, Robert~J Harrison, Wayne~E
  Cascio, and Ana~G Rappold.
\newblock Cardiovascular and cerebrovascular emergency department visits
  associated with wildfire smoke exposure in california in 2015.
\newblock \emph{Journal of the American Heart Association}, 7\penalty0
  (8):\penalty0 e007492, 2018.
\newblock ISSN 2047-9980.

\bibitem[Wu et~al.(2023)Wu, Sverdrup, Mastrandrea, Wara, and Wager]{RN2}
Xiao Wu, Erik Sverdrup, Michael~D Mastrandrea, Michael~W Wara, and Stefan
  Wager.
\newblock Low-intensity fires mitigate the risk of high-intensity wildfires in
  california’s forests.
\newblock \emph{Science Advances}, 9\penalty0 (45):\penalty0 eadi4123, 2023.
\newblock ISSN 2375-2548.

\bibitem[Yadlowsky et~al.(2025)Yadlowsky, Fleming, Shah, Brunskill, and
  Wager]{yadlowsky2025evaluating}
Steve Yadlowsky, Scott Fleming, Nigam Shah, Emma Brunskill, and Stefan Wager.
\newblock Evaluating treatment prioritization rules via rank-weighted average
  treatment effects.
\newblock \emph{Journal of the American Statistical Association}, 120\penalty0
  (549):\penalty0 38--51, 2025.

\bibitem[Zhao et~al.(2022)Zhao, Small, and Ertefaie]{zhao2022selective}
Qingyuan Zhao, Dylan~S Small, and Ashkan Ertefaie.
\newblock Selective inference for effect modification via the lasso.
\newblock \emph{Journal of the Royal Statistical Society Series B: Statistical
  Methodology}, 84\penalty0 (2):\penalty0 382--413, 2022.

\end{thebibliography}

\begin{thebibliography}{9}
\providecommand{\natexlab}[1]{#1}
\providecommand{\url}[1]{\texttt{#1}}
\expandafter\ifx\csname urlstyle\endcsname\relax
  \providecommand{\doi}[1]{doi: #1}\else
  \providecommand{\doi}{doi: \begingroup \urlstyle{rm}\Url}\fi

\bibitem[Athey and Wager(2019)]{SI_athey2019estimating}
Susan Athey and Stefan Wager.
\newblock Estimating treatment effects with causal forests: An application.
\newblock \emph{Observational studies}, 5\penalty0 (2):\penalty0 37--51, 2019.

\bibitem[Breiman(2001)]{SI_breiman2001random}
Leo Breiman.
\newblock Random forests.
\newblock \emph{Machine learning}, 45:\penalty0 5--32, 2001.

\bibitem[Chernozhukov et~al.(2018)Chernozhukov, Chetverikov, Demirer, Duflo,
  Hansen, Newey, and Robins]{SI_chernozhukov2018double}
Victor Chernozhukov, Denis Chetverikov, Mert Demirer, Esther Duflo, Christian
  Hansen, Whitney Newey, and James Robins.
\newblock Double/debiased machine learning for treatment and structural
  parameters, 2018.

\bibitem[Holland(1986)]{SI_holland1986statistics}
Paul~W Holland.
\newblock Statistics and causal inference.
\newblock \emph{Journal of the American statistical Association}, 81\penalty0
  (396):\penalty0 945--960, 1986.

\bibitem[Radcliffe(2007)]{SI_radcliffe2007using}
Nicholas Radcliffe.
\newblock Using control groups to target on predicted lift: Building and
  assessing uplift model.
\newblock \emph{Direct Marketing Analytics Journal}, pages 14--21, 2007.

\bibitem[Robins et~al.(1994)Robins, Rotnitzky, and
  Zhao]{SI_robins1994estimation}
James~M Robins, Andrea Rotnitzky, and Lue~Ping Zhao.
\newblock Estimation of regression coefficients when some regressors are not
  always observed.
\newblock \emph{Journal of the American statistical Association}, 89\penalty0
  (427):\penalty0 846--866, 1994.

\bibitem[Robinson(1988)]{SI_robinson1988root}
Peter~M Robinson.
\newblock Root-n-consistent semiparametric regression.
\newblock \emph{Econometrica: Journal of the Econometric Society}, pages
  931--954, 1988.

\bibitem[van~der Laan et~al.(2011)van~der Laan, Rose, Zheng, and van~der
  Laan]{SI_van2011cross}
Mark~J van~der Laan, Sherri Rose, Wenjing Zheng, and Mark~J van~der Laan.
\newblock Cross-validated targeted minimum-loss-based estimation.
\newblock \emph{Targeted learning: causal inference for observational and
  experimental data}, pages 459--474, 2011.

\bibitem[Yadlowsky et~al.(2025)Yadlowsky, Fleming, Shah, Brunskill, and
  Wager]{SI_yadlowsky2025evaluating}
Steve Yadlowsky, Scott Fleming, Nigam Shah, Emma Brunskill, and Stefan Wager.
\newblock Evaluating treatment prioritization rules via rank-weighted average
  treatment effects.
\newblock \emph{Journal of the American Statistical Association}, 120\penalty0
  (549):\penalty0 38--51, 2025.

\end{thebibliography}
\end{document}